\documentclass[preprint,12pt,authoryear]{elsarticle}

\usepackage{amssymb}
\usepackage{amsmath}
\usepackage{xcolor}
\usepackage{float} 
\usepackage{mathptmx}
\usepackage{etoolbox}
\usepackage{float}
\usepackage{verbatim}
\usepackage{subcaption}
\usepackage{caption}
\usepackage{epsfig}
\usepackage{amsfonts}
\usepackage{comment}
\usepackage{breqn}
\usepackage{graphicx}
\usepackage{hyperref}
\usepackage{ulem}
\usepackage{booktabs}
\usepackage{chngcntr}
\usepackage{svg}

\journal{Nuclear Physics B}

\begin{document}

\begin{frontmatter}



\title{Design and mechanical analysis of the \textsc{Pragya} tokamak vacuum vessel} 


\author[label1]{Ravi Gupta}
\author[label1]{Rahul Babu Koneru}
\author[label1]{Saptarshi Rajan Sarkar}
\author[label2]{Santosh Ansumali}
\author[label3]{Animesh Kuley}
\author[label1]{Roshan George}
\author[label1]{Shaurya Kaushal\corref{cor1}}

\cortext[cor1]{shaurya@pranosfusion.energy}

\affiliation[label1]{
organization={Pranos Fusion Private Limited, Innovation and Development Centre, JNCASR},
city={Bengaluru},
state={Karnataka},
postcode={562162},
country={India}
}

\affiliation[label2]{
organization={Engineering Mechanics Unit, Jawaharlal Nehru Centre for Advanced Scientific Research},
city={Bengaluru},
postcode={560064},
country={India}
}

\affiliation[label3]{
organization={Department of Physics, Indian Institute of Science},
city={Bangalore},
postcode={560012},
country={India}
}

\begin{abstract}

\textsc{Pragya} is a new and India's first privately developed low aspect ratio tokamak
device designed by Pranos Fusion Energy. The tokamak is designed for a
plasma major radius ($R_0$) of about 0.4 m, plasma minor radius ($a$)
greater than 0.18 m, plasma current ($I_p$) of  up to 25 kA and a toroidal
magnetic field ($B_T$) of 0.1 T. The \textsc{Pragya}
vacuum vessel has distinctive features such as a toroidal electric break to
minimize induced eddy current and a double O-ring arrangement to minimize the loss
of vacuum. This paper details the final design of the \textsc{Pragya}
vacuum vessel and presents a comprehensive three-dimensional (3D) finite element model
(FEM) assessment of its structural performance. Evaluations cover the effects of self-weight, atmospheric load and thermal stress arising from in-situ baking. The results confirm that the design satisfies all required safety margins under these combined loading conditions, providing a robust foundation for subsequent plasma operations in this compact tokamak.
\end{abstract}




\end{frontmatter}


\section{Introduction} \label{sec:intro}
An ever-increasing global energy demand, largely met by fossil fuels, is accelerating anthropogenic climate change. Robust low carbon energy alternatives are essential to reduce greenhouse gas emissions and nuclear fusion offers a safe and energy dense alternative to the
existing set of green energy technologies. 
Magnetic-confinement fusion (MCF), with the tokamak as its most extensively studied and experimentally validated configuration, aims to satisfy the Lawson energy-balance requirement by confining a 
$10$ keV plasma long enough that alpha self-heating together with auxiliary heating exceeds losses due to transport and radiation \citep{lawson1957some, ITER_Physics_Expert_Group_on_Confinement_and_Transport_1999}. The current barriers to tokamaks as commercial power sources are primarily around achieving high energy confinement despite microturbulence driven anomalous transport \citep{horton1999drift}; maintaining stable operation against MHD instabilities and disruptions while controlling current and pressure profiles \citep{hender2007mhd}; and realizing a durable plasma–material interface that can simultaneously handle extreme steady state and transient heat fluxes and provide reliable power and particle exhaust (divertor/first-wall performance) \citep{federici2001plasma, herrmann2002overview}. 

Against this multi-constraint background, decades of research has produced major advances in both tokamak physics and the enabling technologies. The
record fusion gain and triple product by JET \citep{keilhacker1999jet}, TFTR
\citep{strachan1997tftr} and JT-60U \citep{kishimoto2005advanced}, discovery of the new operational regime H-mode by ASDEX-U \citep{wagner1982regime},
development of spherical tokamaks such as MAST \citep{darke1995MAST}, NSTX
\citep{ono2001overview}, allowing for a high plasma $\beta $ ($\beta$=plasma
pressure/magnetic pressure) operations in a compact space. 
In parallel, state-of-the-art high-temperature superconducting (HTS) rare earth barium copper oxide (REBCO) magnets have demonstrated large-scale, $\sim$20\,T-class fields, strengthening the case for compact high-field tokamak concepts \citep{hartwig2023sparc, whyte2023experimental}. 
High-performance computing and integrated predictive modelling frameworks increasingly couple microturbulence, transport, pedestal and equilibrium physics, MHD stability and engineering constraints, enabling a more constrained scenario development and design iteration \citep{staebler2022advances, lyons2023flexible, Jenko2025}. Novel plasma control systems based on machine learning are being actively developed for improved flexibility and generalizability \citep{Degrave2022,kerboua2024}.
With ITER and SPARC progressing towards first plasma operations, the field is now positioned to test fusion net-gain relevant regimes with substantially improved physics understanding and technology readiness.

\begin{figure}[hbt!]
    \centering
    \includegraphics[scale=0.8]{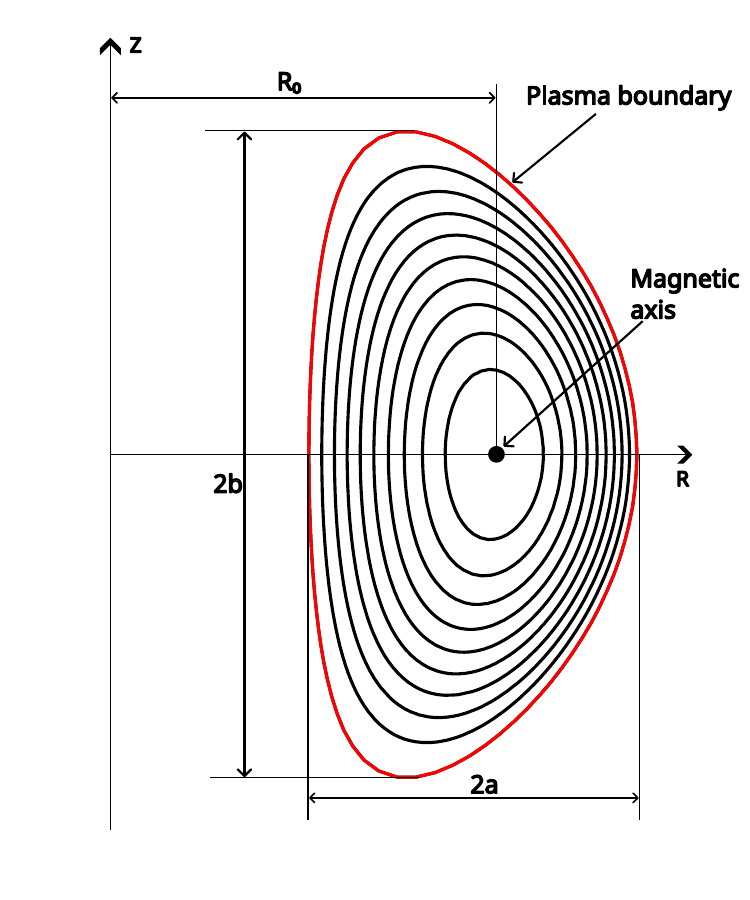}
    \caption{A schematic of the plasma geometry depicting the magnetic contours inside the last closed flux surface (LCFS) which is shown in red. The plasma major radius $R_0$ is defined as the radial distance from the axis of rotation to the plasma magnetic axis. The plasma minor radius $a=(R_{max}-R_{min})/2$ where $R_{max}$ and $R_{min}$ are the maximum and the minimum radial extents of the LCFS respectively. The ellipticity $\kappa=b/a$ and triangularity $\delta=(R_{geo}-R_{up/low})/a$ where $R_{geo}$ is the geometric center of the plasma and $R_{up/low}$ is the radial extent of the uppermost or the lowermost point on the LCFS.}
    \label{fig:plasma_schematic}
\end{figure}

A central design variable in tokamaks is the aspect ratio 
$A$, which strongly couples to MHD stability limits and engineering integration. The aspect ratio is defined as the ratio of plasma major radius $R_0$ to the plasma minor radius $a$. A schematic of the plasma geometry is shown in Figure \ref{fig:plasma_schematic}.
From a magnetohydrodynamic (MHD) stability point of view, a higher plasma
$\beta$ offers better stability against ballooning and kink modes. As the plasma
$\beta$ scales as $1/A$, a smaller $A$ is favoured for stable operations \citep{millerstable1996}. At one end are
the ultra-compact spherical tokamaks (STs) such as PEGASUS-III \citep{sontag2022new}, MAST \citep{darke1995MAST}, NSTX \citep{ono2001overview}, SUNIST \citep{ying2003initial}, LATE \citep{maekawa2005formation}, LTX-$\beta$ \citep{boyle2023extending,banerjee2024investigating}
with $A < 2$.  These machines allow for operations at a high $\beta$, a high
plasma current and a high bootstrap fraction \citep{peng2000physics,
ono2015recent}. However, the lack of space at the high-field side poses a challenge to placing a central solenoid, shielding and blankets. On the other end of the aspect ratio spectrum are EAST ($A=4.25$) \citep{wu2007overview}, SST-1 ($A=5.5$) \citep{deshpande1997sst}, WEST ($A>5.6$) \citep{bucalossi2022operating} which are not constrained for space but are prone to MHD instabilities. The remainder of the aspect ratio range has tokamaks such as DIII-D ($A=2.5$) \citep{luxon2002design}, JT-60SA ($A=2.5$) \citep{tomarchio20173}, ADITYA ($A=3.0$) \citep{bhatt_bora_buch_and_others_1989}, Alcator C-Mod ($A=3.09$) \citep{marmar2007alcator}, SPARC ($A=3.24$) \citep{creely2020overview}, KSTAR ($A=3.6$) \citep{lee2001design} which offer exploration of various physics and subsystems for ITER and DEMO-like machines \citep{tomarchio20173}. 
Wong and co-authors \citep{cpcwong_2002} in their analysis observed that for an elongation of 2, the minimum cost of electricity was obtained for $A \in [2,3]$  and $A \sim 2$ for a superconducting (SC) reactor and a normal conducting (NC) reactor respectively. The analysis explored the combined effect of aspect ratio ($1.2-6.0$), elongation ($1.5,3.0$) and MHD stability on the design of NC and SC fusion reactors. From an economics standpoint, along with the trade-off between available space inside a tokamak and MHD stability, the low aspect ratio tokamaks are very promising in the context of a pilot fusion power plant.
Within this landscape, \textsc{Pragya} is being developed as India's first privately developed low aspect ratio tokamak designed by
Pranos Fusion Energy. \textsc{Pragya} is designed for a plasma major radius $R_0$ of 0.4 m and minor radius $a$
of 0.18 m which gives $A \sim 2.2$ with a planned
operational plasma current $I_p$ of about 25 kA. The magnetic system consists of a central solenoid along
with the toroidal and poloidal field coils. The toroidal field coils are designed to
produce a magnetic field $B_T$ of strength 0.1 T at $R_0$. The plasma current will be
driven by a central solenoid while a 6 kW microwave source operating at 2.45 GHz
will be used for pre-ionization and plasma startup. The target vacuum pressure is $10^{-6}$ mbar. This is a small compact tokamak designed as a precursor to
a larger tokamak with scientific exploration and development of critical human resources as a core
objective. In particular, \textsc{Pragya} is designed to address the following
objectives:
\begin{enumerate}[i.]
    \item accommodate plasma with an elongation between 1.5 and 2.2 and a triangularity between 0.0 and 0.4 
    \item test bed for superconductor magnets, auxillary heating, solenoid-free start-up and current drive, novel plasma control systems for future large-scale fusion device
    \item investigate the magnetohydrodynamic (MHD) stability of plasma
    \item test remote handling.
\end{enumerate}

The vacuum vessel is a crucial component of tokamak maintaining a very high
vacuum necessary for plasma generation, confinement and shaping. In addition to
the vacuum load and the self-weight, the vacuum vessel is also subjected to
thermal load during baking operations, electromagnetic loads from the magnetic
system and also large asymmetric loads encountered during plasma disruption
events. Additionally, a vacuum vessel provides support for plasma facing
components such as limiters and divertors along with diagnostics such as flux
loops and Rogowski coils. Furthermore, the vacuum vessel also acts as a support
structure for the toroidal field (TF) and the poloidal field (PF) coils. The design of the vessel should ensure requisite level of
vacuum pressure and structural integrity against the aforementioned loads all
the while sparing enough room for diagnostics. 

The primary focus of this paper is the design of \textsc{Pragya} tokamak's vacuum vessel.
The vessel is designed to strictly adhere to the safe structural and thermal
loads encountered during the tokamak operation. Detailed simulations are
conducted to assess the structural integrity of vacuum vessel. The remainder of
the paper is organized as follows.  The mechanical design of the vacuum vessel
and a detailed description of its various components are presented in Section
\ref{sec:mech_design}. The results from the thermo-mechanical stress analysis
are discussed in Section \ref{sec:stress_analysis} wherein results from 3D FEA
analysis of mechanical and thermal loads on the vacuum vessel are discussed.
Finally, the conclusions are drawn and a future outlook of \textsc{Pragya} is provided in
Section \ref{sec:conclusions}

\section{Mechanical design of the vacuum vessel} \label{sec:mech_design}
\subsection{Description of the vacuum vessel} \label{ssec:vv_description}
\begin{figure}[hbt!]
    \centering
    \includegraphics[width=0.9 \linewidth]{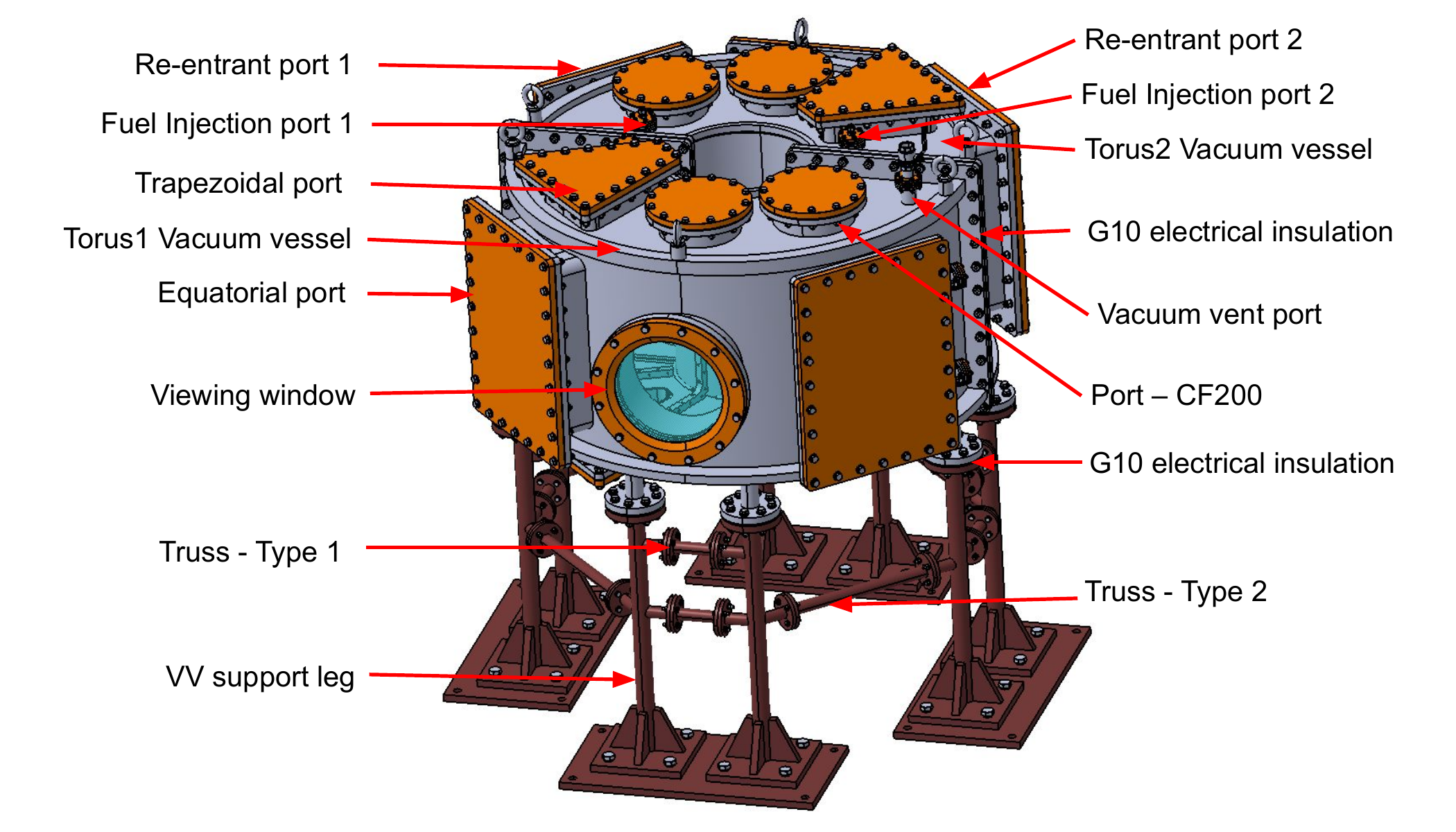}
    \caption{3D CAD model of the \textsc{Pragya} vacuum vessel (VV) assembly.}
    \label{fig:PRAGYA_3D_CAD}
\end{figure}

\textsc{Pragya} tokamak is being designed for a plasma major radius of 0.4 m, a plasma
minor radius of 0.13 m - 0.18 m and a plasma current of upto 25 kA. The resulting
aspect ratio is between 2.2 and 3.1. The key target operational parameters of \textsc{Pragya} are
listed in Table \ref{tab:PlasmaDimensions}. The bounds of the operational limits are subject to safety factor on axis $q_0$ and on the edge $q_{95}$ that allow for a stable plasma operation. For a given geometry of the plasma, $q_0$ and $q_{95}$ are computed using the fixed-boundary Grad-Shafranov solver in TokaMaker \citep{hansen2024109111}. These values are listed for the minimum and maximum values of $A$ and $\delta$ in Table \ref{tab:safetyfactor}. A short description of the Grad-Shafranov equation is provided in \ref{Appn:C}. Based on the aforementioned
requirements and budgetary constraints, a vacuum vessel of width 400
mm and a height of 600 mm is designed. The wall thickness of the vacuum vessel
is 6 mm on all sides. This particular choice of the wall thickness was result of
an optimization procedure set to minimize the maximum stress on the vessel and
is discussed in Section \ref{sec:stress_analysis}. The resulting inner diameter
(ID) and the outer diameter (OD) of the vessel are $388$ mm and $1212$ mm
respectively with a total interior volume of $0.8\ \mathrm{m^3}$. The 3D
computer aided design (CAD) model of the vacuum vessel assembly is shown in
Figure \ref{fig:PRAGYA_3D_CAD}. To provide electrical insulation, the vacuum
vessel is divided into two sub-tori with an electrical isolation material, G10,
sandwiched between them. The electrical break is discussed in more detail in
Section \ref{ssec:electrical_break}. The inner diameter (ID), outer diameter
(OD) and wall thickness of the torus are $388$ mm, $1212$ mm, and $6$ mm,
respectively. The total interior volume (or, vacuum volume) of the vessel is
$0.8\ \mathrm{m^3}$. A combination of circular, rectangular and trapezoidal
ports, 22 in total, are present on the vacuum vessel. The description of various
ports and their utility is presented in Section \ref{sssec:ports}. The vacuum
vessel is made of stainless steel 304L. 

\begin{table}[H]
    \centering
    \begin{tabular}{l l}
       \hline
       Geometric feature  & Value  \\
       \hline
       Major radius, $R_0$     & $0.4$ m  \\
       Minor radius, $a$       & $0.13\ \textrm{m} - 0.18$ m \\
       Aspect ratio, $A$       & $ 2.22 - 3.15$ \\
       Elongation,   $\kappa$  & $1.55-2.20$\\
       Triangularity, $\delta$ & $0.0 - 0.4$ \\
       \hline  
    \end{tabular}
    \caption{Geometric features of the plasma in \textsc{Pragya} vacuum vessel}
    \label{tab:PlasmaDimensions}
\end{table}

\begin{table}[!htbp]
\centering
\begin{tabular}{*5c}
\toprule
$A$ &  \multicolumn{2}{c}{$\delta=0$} & \multicolumn{2}{c}{$\delta=0.4$}\\
\midrule
{}   & $q_0$   & $q_{95}$    & $q_0$   & $q_{95}$\\
\cline{2-5}
2.22   &  1.64 & 3.60   & 1.47  & 4.30 \\
3.15   &  1.59 & 2.68   & 1.37  & 3.10 \\
\bottomrule
\end{tabular}
\caption{The safety factor on the axis $q_0$ and at the edge $q_{95}$ for the extreme values of the aspect ratio $A$ and triangularity $\delta$ computed from the fixed-boundary Grad-Shafranov solver TokaMaker.}
\label{tab:safetyfactor}
\end{table}

\subsection{Design of the vacuum vessel} \label{ssec:vv_design}

\subsubsection{Ports} \label{sssec:ports}
\begin{figure}[hbt!]
    \centering
    \includegraphics[width=0.8 \linewidth]{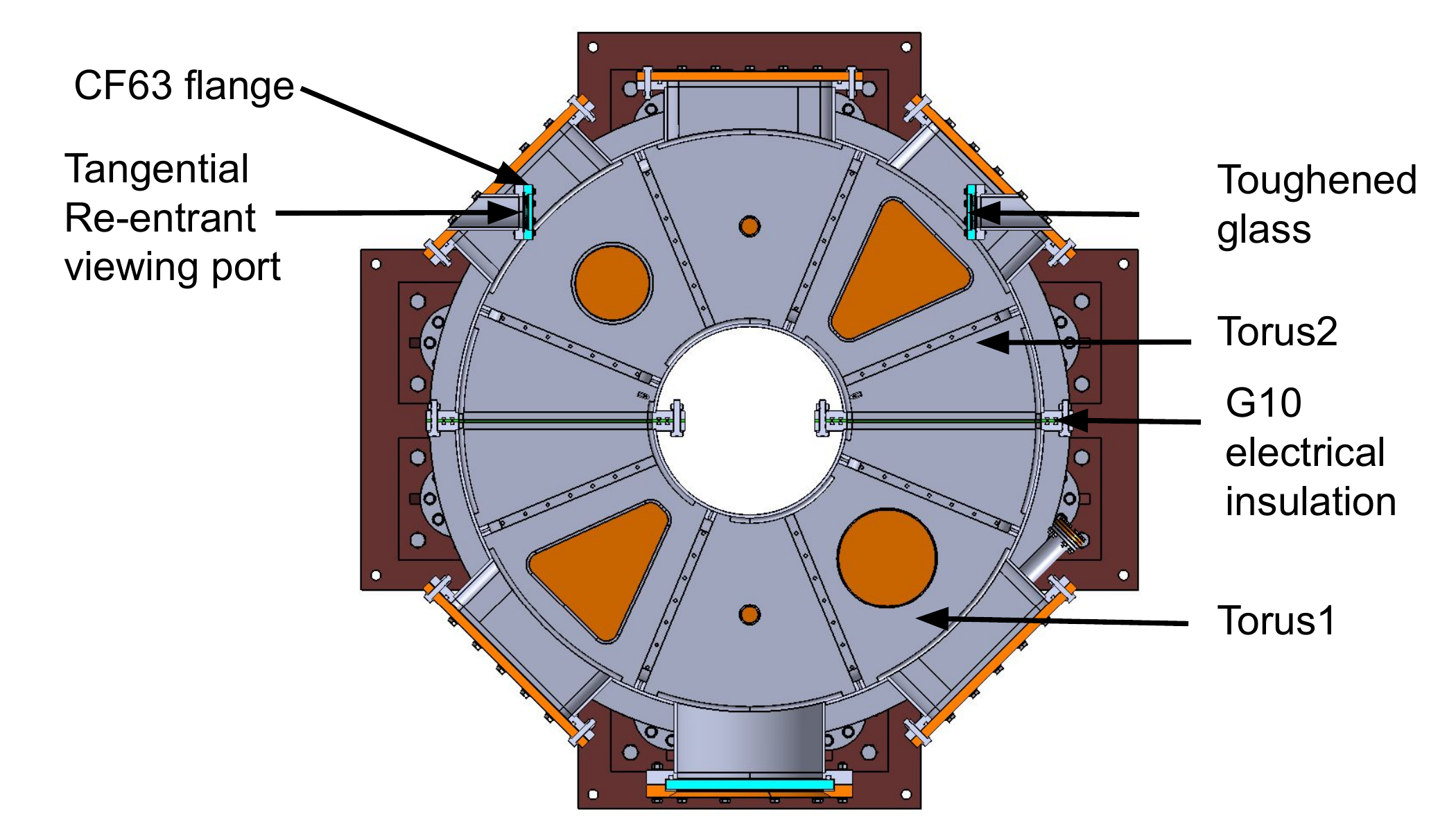}
    \caption{A cut-section view of the 3D CAD model of tangential viewing port in the \textsc{Pragya} vacuum vessel assembly.}
    \label{fig:TangentialViewingPort_3D_CAD}
\end{figure}

Each of the two sub-tori house multiple ports along the equatorial, top, and
bottom sides. These ports will be used for diagnostics,  connecting the pumping
system, fuel inlet, and microwave heating source. The layout of the ports on
the vacuum vessel is shown in Figure \ref{fig:PRAGYA_3D_CAD} and Figure
\ref{fig:TangentialViewingPort_3D_CAD} with a brief description of all the ports
provided below.

\paragraph{Viewing port}
A circular viewing port (diameter $= 300$ mm) is provided in the equatorial direction (shown in Figure \ref{fig:PRAGYA_3D_CAD}, \ref{fig:TangentialViewingPort_3D_CAD}). The viewing port has a $20$ mm thick toughened glass with O-ring sealing provided on the vacuum and atmosphere sides. This port provides visual access to high-speed cameras or for capturing plasma emissions using photomultiplier tubes (or photodiodes).

\paragraph{Trapezoidal-shaped port}
A pair of trapezoidal ports are placed on the top and bottom sides of the vacuum vessel for diagnostics such as microwave interferometry. The trapezoidal shape allows a broader access in the radial direction and a closer reach on the inboard side of the vacuum vessel. 

\paragraph{Diameter Nominal (DN) $200$ ConFlat (CF) ports}
Four DN $200$ CF ports are provided on the top and one DN $200$ CF port is provided bottom side of the vessel. The DN $200$ ports placed on the top face of the vacuum vessel will be used for different diagnostics such as Langmuir probe. The bottom DN $200$ port serves a dual purpose as a backup port for connecting the vacuum pump and as a port for diagnostics. 

\begin{figure}[hbt!]
    \centering
    \includegraphics[width=0.8 \linewidth]{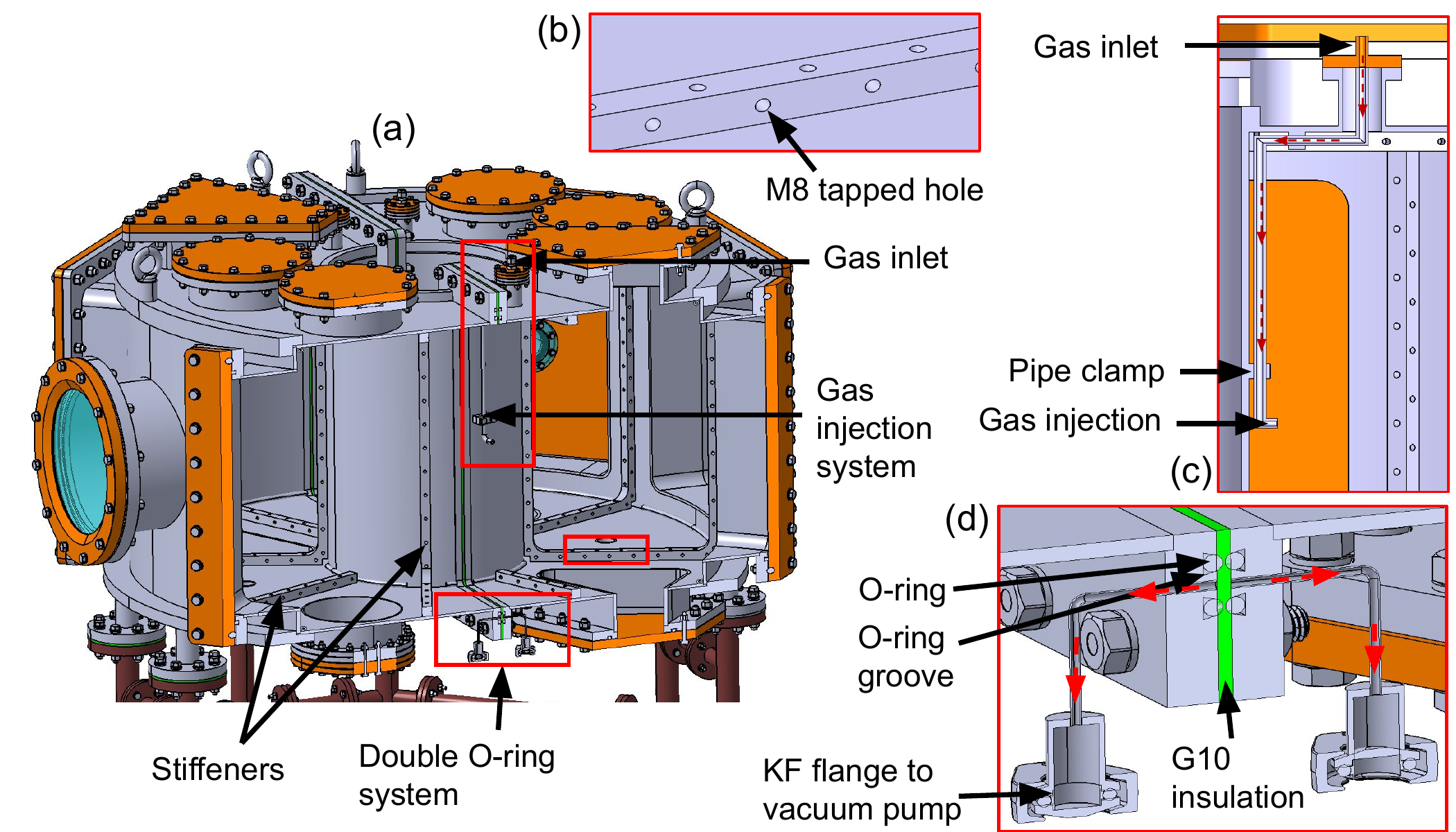}
    \caption{(a) A cut-section view of the 3D CAD model of the vacuum vessel assembly; zoomed-in descriptive view of (b) stiffener, (c) gas inlet system (dotted red arrows represent the direction of incoming gas flow), (d) double O-ring vacuum sealant assembly (dotted magenta arrows represent the direction of flow due to vacuum produced between the two O-ring arrangement).}
    \label{fig:zoom_stiffener_GasInlet_doubleO_ring}
\end{figure}

\paragraph{Gas injection port}
Two DN $40$ ports are placed on the top face of the vessel for gas injection (shown in Figure \ref{fig:PRAGYA_3D_CAD}). A DN $3$ SS304L ($OD$ = $10.3$ mm) pipe runs from each DN $40$ CF port into the chamber while hugging along the wall (as shown in Figure \ref{fig:zoom_stiffener_GasInlet_doubleO_ring}a, \ref{fig:zoom_stiffener_GasInlet_doubleO_ring}c). These SS304L pipes run till the midplane and the gas is injected in the radial direction. To avoid vibrations of the inlet pipe during operations, they are held to the vacuum vessel using a clamp arrangement.  

\paragraph{Equatorial rectangular port}
A total of five identical rectangular ports are provided on the outboard side of
the vessel along the toroidal direction. These ports are $340$ mm in width and
$500$ mm in height. These ports have multiple utilities such as microwave
injection, visualization and diagnostics. Figure
\ref{fig:TangentialViewingPort_3D_CAD} shows tangential entry port assembly in
the \textsc{Pragya} vessel. The viewing port has toughened glass mounted on a DN
$63$ CF flange. 
The radial coverage of the tangential entry viewing port is $63$
mm ranging from $384$ mm till $447$ mm. 

\paragraph{Turbomolecular pump (TMP) port}
A DN $160$ CF port is placed on the bottom side of the vacuum vessel to connect a Pfeiffer HiPace 700 H turbomolecular pump (TMP) (as shown in Figure \ref{fig:TangentialViewingPort_3D_CAD}). The TMP is not directly mounted on the designated TMP port to avoid a large magnetic field protruding inside the TMP. The safe limit of external magnetic field for the TMP is $6$ mT (provided in the operating instructions manual of Pfeiffer HiPace 700 \footnote{\href{https://www.lesker.com/newweb/vacuum_pumps/pdf/manuals/hipace700-manual.pdf}{Pfeiffer HiPace 700 operating instructions manual}}). Therefore, the TMP is placed $2$ m away from the centerline of the vessel via an L-shaped bellow connected to the TMP port on the vessel. Based on the internal surface area ($4.23\ \mathrm{m^2}$) and the total volume ($0.8\ \mathrm{m^3}$) of the vacuum vessel along with the distance of the TMP from the vessel, reaching a vacuum of order $10^{-6}$ mbar will take approximately 20 hours. This can be reduced to 8 hours with the introduction of a booster pump.  

\paragraph{Roughing pump port}
Two DN $40$ ports are placed on the bottom side of the vacuum vessel to connect Pfeiffer DuoVane 18 rotary vane pump (as shown in Figure \ref{fig:TangentialViewingPort_3D_CAD}). The details of the vacuum pumping are provided in Section \ref{sec:vacuum_pump_system}.


\paragraph{Vacuum vent port}
A DN $40$ port is placed on the top side of the vacuum vessel for venting out the vacuum inside the vessel (as shown in Figure \ref{fig:PRAGYA_3D_CAD}). For activities such as periodic maintenance and setting-up new diagnostics, the vessel needs to be purged and brought to atmospheric condition. Purging will be done by first ensuring all the pumps and gate valves are closed, and later injecting dry $\mathrm{N_2}$ gas into the vessel.   

\subsection{Electrical break} \label{ssec:electrical_break}
The characteristic input current to the electrical coils responsible for magnetic field and plasma current involves current ramp-up and ramp-down stages. During the ramp-up or ramp-down stage, the change in magnetic flux leads to induced eddy current in the vacuum vessel walls. The induced eddy currents generate magnetic fields that interact with the plasma and affects its stability. 
In elongated tokamak plasmas, vertical stability is inherently unstable  as any small vertical displacements grow exponentially without control \citep{fukuyama1975positional}. 
Eddy currents in the vacuum vessel create up-down asymmetric radial fields that can either passively stabilize (reduce growth rate) or destabilize the plasma (worsen vertical instability and accelerate VDEs), depending on vessel geometry, time scale, and current mode.
Furthermore in a toroidally continuous vacuum vessel, a major part of the Ohmic flux swing is lost in driving the current through the walls. This could be problematic in small machines where the available volt-seconds are minimal to start with. To minimize these effects, the vessel is divided into two sub-tori along the toroidal direction. A similar approach has been followed in existing tokamaks such as ADITYA-U
\citep{jadeja2017aditya}, SUNIST \citep{li2015measurement} and ASDEX \citep{finkelmeyer1979assembly}. An electrical insulation material G10, is sandwiched between the two sub-tori to provide electrical insulation in the toroidal direction. Figures \ref{fig:PRAGYA_3D_CAD} and \ref{fig:zoom_stiffener_GasInlet_doubleO_ring}d show the placement of G10 material in between the two torus. Additional G10 insulation is provided between the vessel and the support structure, as shown in Figure \ref{fig:PRAGYA_3D_CAD}, to further minimize eddy currents.

To ensure a good vacuum sealing at the junction where the two sub-tori are placed together (with G10 in between them), the flanges of the two sub-tori are provided with a double O-ring arrangement as shown in Figures \ref{fig:zoom_stiffener_GasInlet_doubleO_ring}a and
\ref{fig:zoom_stiffener_GasInlet_doubleO_ring}d. A series of two additional O-rings are placed along the poloidal direction and the space between them is evacuated using a roughing pump. The vacuum in this space will be maintained between $10^{-2}$ and $10^{-3}$ mbar. 
This will ensure that any atmospheric gas passing through the outer O-ring will be evacuated in the space between the two O-rings.  

\subsection{Stiffeners} \label{Sec:MechDesign_Stiffener}
The vacuum facing side of each tori houses ribs running along the poloidal direction. Figures
\ref{fig:zoom_stiffener_GasInlet_doubleO_ring}a, \ref{fig:zoom_stiffener_GasInlet_doubleO_ring}b show the stiffeners in the vacuum vessel
assembly. Although the addition of these ribs reduces the available volume for plasma, these ribs act as stiffeners providing 
structural rigidity to the vessel. These ribs have a cross-section $20 \times 20$ $\mathrm{mm^2}$. A total of eight ribs are present in
the toroidal direction. Eight stiffeners are added on the top plate, bottom plate and outer cylinder. All these stiffeners are present 
on the inner faces. Four stiffeners are present on the inner face of inner cylinder. Furthermore, these ribs are lined with standard M8 tapped holes on three sides which will be used to mount 
limiter or diagnostic equipment inside the vacuum vessel. These tappings are $50$ mm apart and are staggered on the adjacent faces. Figure \ref{fig:Appn:WithAndWithoutStiffener_von_mises_stress} in \ref{Appn:B} shows a comparison of stress induced on a vacuum vessel with and without stiffeners. With the addition of stiffeners, a 6 to 7 times reduction in the maximum stress is observed. 

\subsection{Vacuum vessel support structure}
The vacuum vessel rests on a support structure that consists of $8$ legs ($4$ on each
sub-torus). The legs are $0.75$ m tall which puts the center of
the circular viewing port $1.2$ m above the ground. These legs are connected to
the vessel at $620$ mm away from the central axis to 
provide sufficient space at the bottom side of the vessel connecting vacuum
pumps and diagnostics. These legs rest on base plate which 
will be grouted to the floor of the facility. The support legs are designed to
hold the weight of the vessel without undergoing buckling. 
The details of linear buckling analysis performed on the support legs are
provided in the Section \ref{ssec:linearBuckling_supportLeg}. The legs are
connected through trusses in the toroidal direction to give stability to the vessel
against the lateral forces or moments acting onto it.   

\subsection{Material for construction}
A common choice for the tokamak vacuum vessels is stainless steel due to its
non-magnetic nature, good machinability characteristics, lower cost and
corrosion resistance. The two widely used grades of stainless steel in the construction of tokamak vacuum vessels are 304 (L)
\citep{ying2003initial, song2006design, jadeja2017aditya,  sontag2022new} and
316 (L) \citep{sakurai2009design, mardani2012design, chung2013design,
mancini2021mechanical, huang2024electromagnetic} which differ by their corrosion
resistance properties and the amount of carbon present in them. The vacuum vessel, ports,
blanks of the ports, support structures are all made of SS304L. SS304L is chosen
due to its lower magnetic permeability, high mechanical strength, corrosion
resistance, low electrical conductivity, ease of availability, established
fabrication/machining processes and also the relatively low cost compared to SS316.

In summary, the key characteristics of the
\textsc{Pragya} vacuum vessel assembly are listed below.

\begin{enumerate}[i.]
    \item The vacuum vessel is electrically isolated from the induced eddy current in the toroidal direction by dividing it into two sections.
    \item To reduce leaks, a double O-ring arrangement is designed on the flanges and the space between the two sets of O-rings is evacuated using a vacuum pump.
    \item Multiple ports are present on the equatorial, top and bottom sides of the vacuum vessel.
    \item Use of ribs (or, stiffeners) to provide additional structural strength to the vacuum vessel which allows for mounting limiter or diagnostics inside the vacuum vessel.
\end{enumerate}

\subsection{Vacuum pumping system} \label{sec:vacuum_pump_system}
The \textsc{Pragya} vacuum vessel is designed for a maximum vacuum of $10^{-6}$ mbar. To generate a vacuum of this order, a series of two pumps,
a roughing pump (Pfeiffer Duovane 18 rotary vane pump\footnote{\href{https://www.pfeiffer-vacuum.com/fi/assets/121014/1/PD0118BEN_B.pdf}{Pfeiffer Duovane 18 rotary vane pump data sheet}}) of capacity 3.9 l/s and a turbomolecular pump (TMP) 
(Pfeiffer HiPace 700 H\footnote{\href{https://www.pfeiffer-vacuum.com/global/assets/132740/1/Product\%20Leaflet\%20HiPace\%20700\%20H\%20Global\%20EN.pdf}{Pfeiffer HiPace 700 H data sheet}}) of capacity 665 l/s, are used. The TMP is connected to the bottom face of the vessel in the designated DN $160$ CF port. The CF has a knife edge and an OFHC copper gasket to provide a vacuum seal. A splinter plate and a manual/electropneumatic gate valve is provided in between the DN $160$ port and the TMP. The splinter plate prevents any unwanted material dropping into the TMP. The gate valve provides control to connect/disconnect the TMP from the vessel. The RVP is connected to the DN $40$ CF port at the bottom face of the vacuum vessel.

\section{Thermo-mechanical stress analysis} \label{sec:stress_analysis}

During operation, the vacuum vessel will be subjected to different loads, 

\begin{enumerate}[i.]
    \item pressure differential due to vacuum inside the vacuum vessel and atmospheric pressure outside the vacuum vessel, 
    \item weight of the vessel,
    \item thermal load during baking and
    \item electromagnetic load.
\end{enumerate}

Different features of the vacuum vessel such as the wall thickness, stiffener locations and
dimensions, radius of curvature of the edges of the ports, are designed such
that the stress induced in the vacuum vessel due to all these loads combined is within the
material's yield strength ($170$ MPa for SS304L). 

Finite element analysis (FEA) is performed on the entire vacuum vessel assembly
(including ports and support structure) to compute the stresses induced due to
the vacuum, self-weight and thermal loads using COMSOL Multiphysics $6.3$. The governing equations pertinent to the results in this paper are presented in \ref{Appn:C}.
Table \ref{tab:SS304_properties} shows the mechanical and thermal and properties of stainless steel SS304L used during the simulations \citep{ASME_BPVC_II_D_2019}. The values at $20\ ^\circ \mathrm{C}$ are used during simulations at room temperature conditions, while the values at $150\ ^\circ \mathrm{C}$ are used for simulations at baking conditions. 


\begin{table}[!htbp]
\centering
\begin{tabular}{lcc}
\toprule
Property &  \multicolumn{2}{c}{Value at} \\
{}   & $20\  ^\circ \mathrm{C}$ & $150\  ^\circ \mathrm{C}$  \\
\midrule
        Density, $\rho$ ($\mathrm{kg/m^3}$) & $8030$  & $8030$ \\
        Young's modulus (GPa) & $195$ & $186$ \\
        Poisson's ratio & $0.31$ & $0.31$\\
        Heat capacity at constant pressure, $C_p$ $(\mathrm{J/(kg \cdot K)})$ & $472$ & $510$ \\
        Thermal conductivity, $k$ $(\mathrm{W/(m \cdot K)})$ & 14.8 & 17.0  \\
        Coefficient of thermal expansion, $\alpha$ ($\times 10^{-6} \mathrm{1/K}$) & $15.3$ & $16.6$ \\
\bottomrule
\end{tabular}
    \caption{Mechanical, thermal and electrical properties of stainless steel SS304L used in the COMSOL simulations. The heat capacity at constant pressure, $C_p=k/(\rho \alpha)$.}
    \label{tab:SS304_properties}
\end{table}

\subsection{Grid Refinement Study} \label{ssec:grid_refinement}
\begin{figure}[hbt!]
    \centering
    \includegraphics[width=1.0\textwidth]{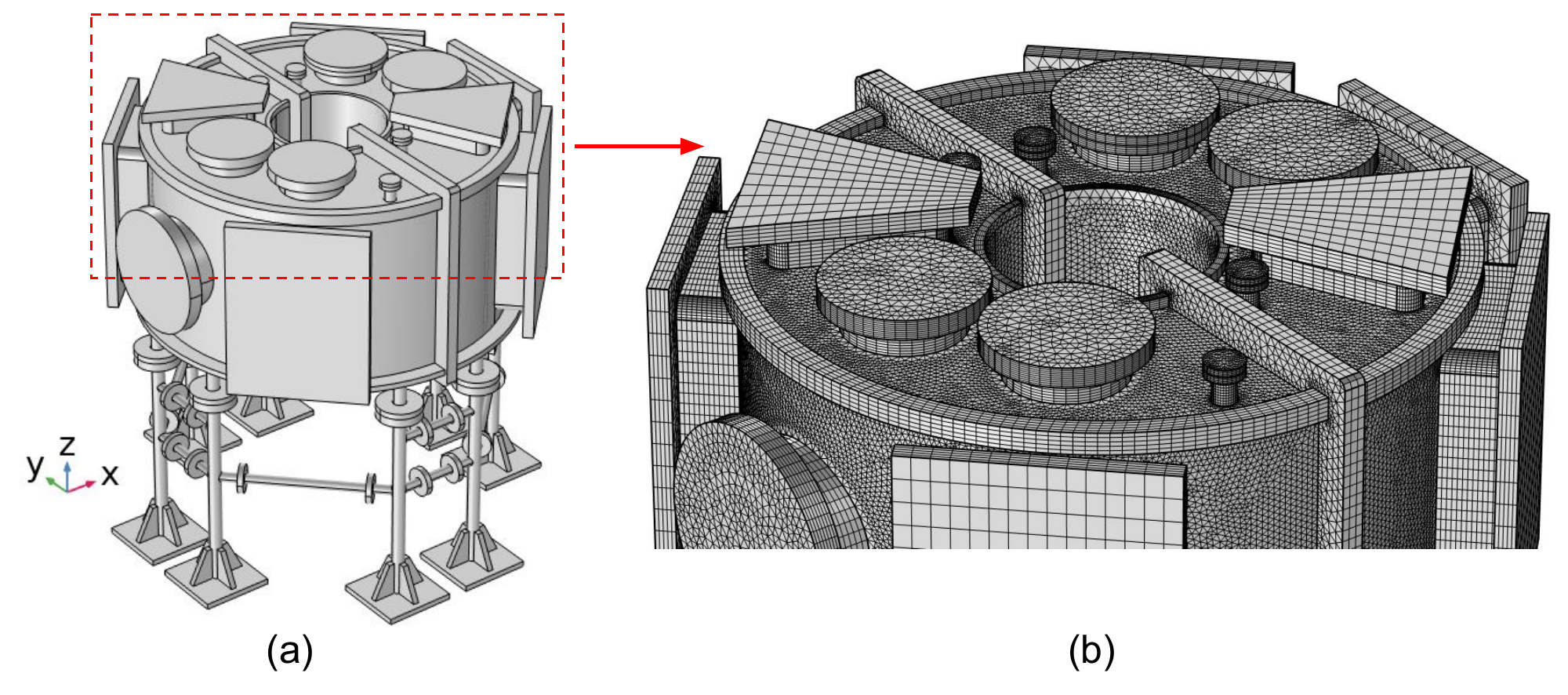}
    \caption{(a) 3D CAD model and (b) mesh used for the simulations.}
    \label{fig:Simulation_CAD_mesh}
\end{figure}

\begin{figure}[hbt!]
    \centering
    \begin{subfigure}{0.49\textwidth}
    \includegraphics[width=\textwidth]{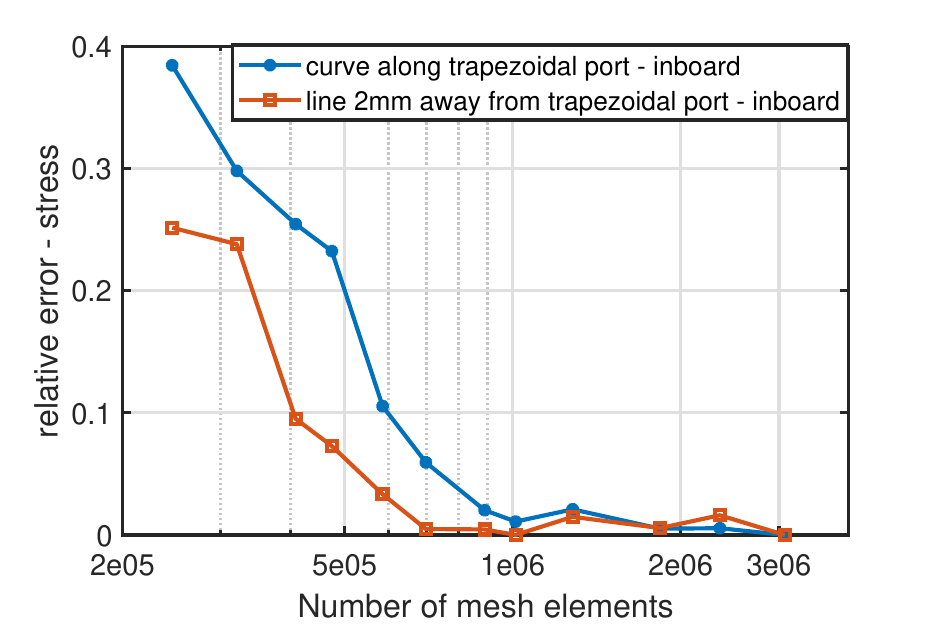}
    \caption{}
    \label{fig:MeshConvergence_PeakStress}
    \end{subfigure}
    \begin{subfigure}{0.49\textwidth}
    \includegraphics[width=\textwidth]{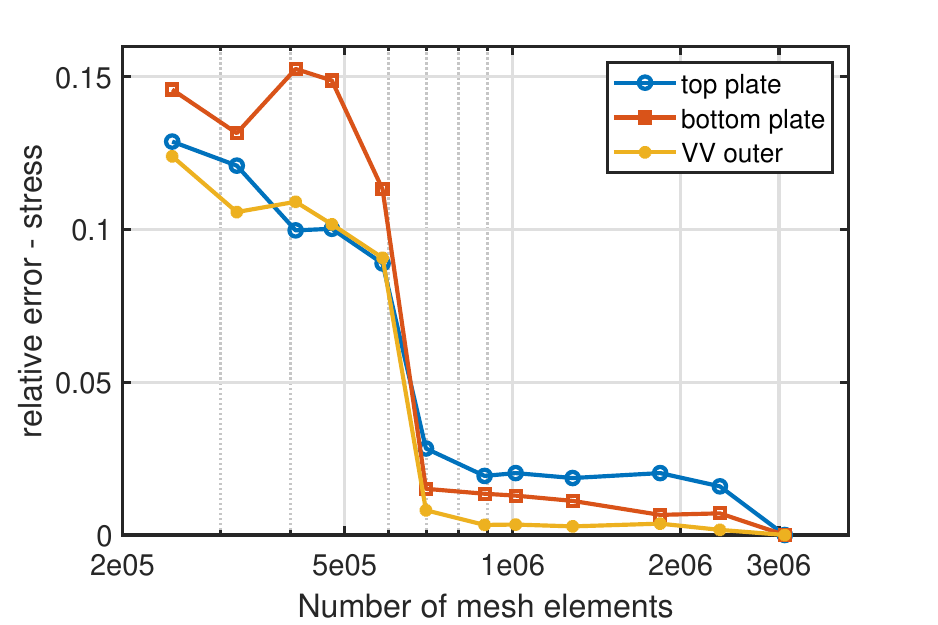}
    \caption{}
    \label{fig:MeshConvergence_SurfaceAvgStress}
    \end{subfigure}
    \caption{Semilogx plots of relative error in (a) von Mises stress averaged along a curve near the
    top trapezoidal port on the inboard side and (b) surface averaged von Mises stress
    averaged along top plate, bottom plate and vacuum vessel outer cylindrical surface.}
    \label{fig:MeshConvergence}
\end{figure}

A grid refinement study is performed to optimize the number of mesh elements
used for the simulations. Figure \ref{fig:Simulation_CAD_mesh}
shows the 3D CAD model and the mesh used for the simulation. At multiple places
in the geometry, quadrilateral mesh elements are used as they provide uniform
and better control of spatial distribution of mesh elements. The small
features of the vacuum vessel assembly such as radius of curvature of the ports,
stiffeners, port thickness, blank thickness, are locally resolved as shown in
Figure \ref{fig:Simulation_CAD_mesh}c to accurately represent the stress and
deformation. To perform the grid convergence study, global and local mesh
element size were changed. For the tetrahedral elements, the global mesh
elements size was modified while for the quadrilateral elements, the local mesh
element size was modified. Two parameters, surface-averaged von Mises stress
and line-averaged von Mises stress are the quantities of interest in this study.
The surface averaged von Mises stress is monitored at three locations - top
plate, bottom plate and outboard cylinder. At certain locations where the
changes in geometry are not smooth enough, a localization (in one or two mesh elements) of stress is observed. In this work, the locations where the ports meet the vessel especially on the inboard side of the trapezoidal ports were observed to be the regions of high stress localization. This is a well-known numerical artefact which can be
resolved with a finer mesh around the problem zone or by analyzing the data a
few mesh elements away from the problem zone. In this work, this issue is
remedied by computing the line-averaged von Mises stress at the inboard side of
the trapezoidal port. In all the simulations, the quantities of interest
converge to a residual error of the order of $10^{-9}$. The effect of mesh
resolution on these quantities is shown in Figure
\ref{fig:MeshConvergence_PeakStress} and Figure
\ref{fig:MeshConvergence_SurfaceAvgStress}. In the absence of a known solution,
the error is measured relative to the solution on the finest grid. These figures
are semilog plots with the abscissa (or number of mesh elements) in the log
scale. As expected, the relative error reduces with an increase in number of
mesh elements or a reduction in the mesh volume. The reduction is quite steep up
to about $700$k elements after which the error plateaus. Therefore, the simulations discussed in subsequent sections is performed on the mesh with $1012$k number of mesh elements. This mesh is shown in the
Figure \ref{fig:Simulation_CAD_mesh}(b-c).  

\subsection{Vacuum vessel wall thickness} \label{ssec:vv_wall_thickness}
\begin{figure}[hbt!]
    \centering
    \includegraphics[width=0.6 \linewidth]{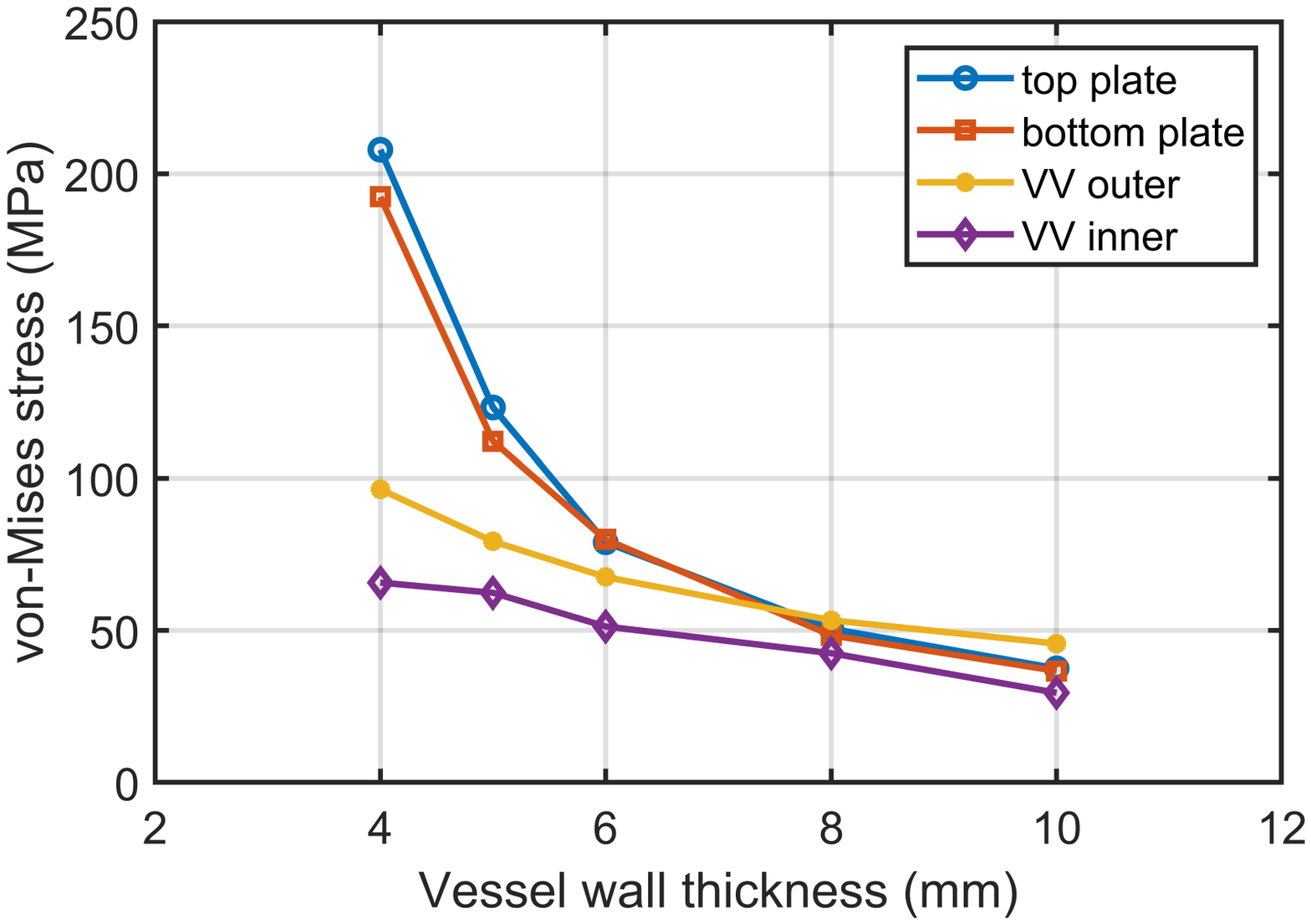}
    \caption{Wall thickness convergence study: Variation of peak von Mises stress at different surfaces of vacuum vessel vs vessel wall thickness.}
    \label{fig:convergence_wallThickness}
\end{figure}

The effect of wall thickness over the maximum stress induced in the vacuum
vessel is evaluated first to narrow down the wall thickness. A convergence study
is performed to evaluate the optimum wall thickness for the vessel. The study is
performed considering the pressure difference due to vacuum and atmosphere, and
self weight of the vessel. Maximum von Mises stress induced on different faces
(top plate, bottom plate, inner cylinder and outer cylinder) of the vessel is
monitored. The geometry and the corresponding mesh for this study is shown in
\ref{Appn:A}. Figure \ref{fig:convergence_wallThickness} shows the
variation of maximum von Mises stress induced at these surfaces with respect to
different wall thickness. Two observations are made: first,
the maximum von Mises stress induced on the different faces of the vessel
increases with a decrease in wall thickness; and second, for a given wall thickness,
the maximum von Mises stress is observed on the top plate except at higher wall
thickness of $8$ and $10$ mm. Therefore, we consider the maximum stress induced
on the top/bottom plate for the optimum wall thickness evaluation. It is
observed that the maximum stress induced on a $4$ mm thick vacuum vessel is approximately
$200$ MPa which is already quite close to the yield limit of SS304L. Increasing
the wall thickness to $5$ mm, $6$ mm and $8$ mm, induces a maximum stress of
$123$ MPa, $79$ MPa and $51$ MPa respectively. While the stress is considerably lower for both the 6 mm and the 8 mm thick wall, a 6 mm thick wall is chosen from a material volume and cost point of view.

\subsection{Stress due to vacuum and self-weight of the vacuum vessel}
To ascertain the structural integrity of the vacuum vessel against a
differential pressure load due to vacuum, all the vacuum facing surfaces are set
to a pressure of $10^{-5}$ Pa, while the atmosphere facing surfaces are set to a
pressure  of $101325$ Pa. The base of the support legs are set to a fixed
boundary condition with gravity acting along the negative $z$ direction.

\begin{figure}[hbt!]
    \centering
        \includegraphics[width=0.8 \linewidth]{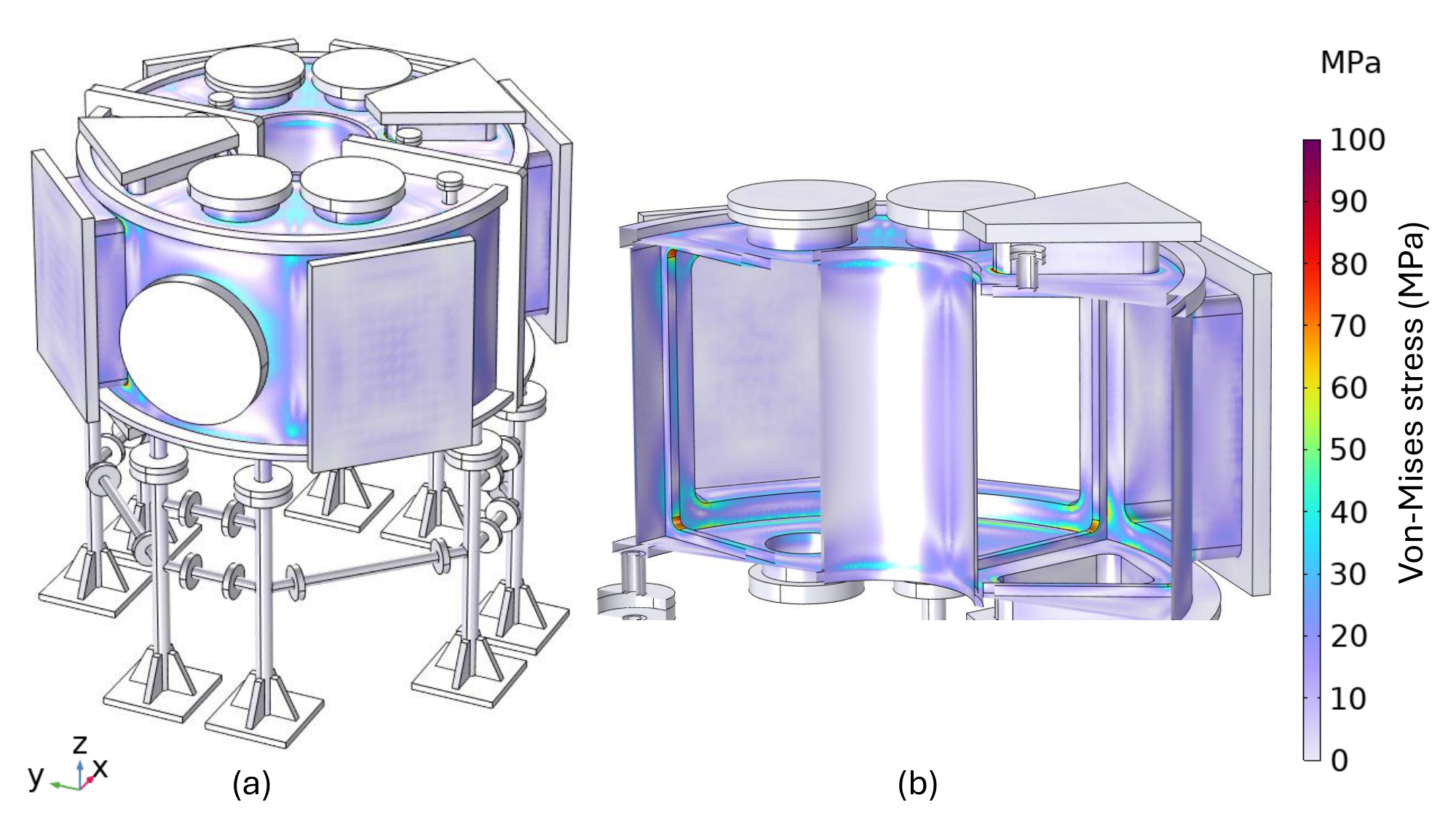}
    \caption{Spatial distribution of stress (von Mises) on the vacuum vessel due to vacuum and self-weight.}
    \label{fig:Pressure_SelfWt_von_mises}
\end{figure}

The resulting von Mises stress on the vessel due to the imposed pressure
difference and self-weight of the vessel is shown in Figure
\ref{fig:Pressure_SelfWt_von_mises}. It is observed that the maximum stress
induced on the vessel is around $110$ MPa observed at two locations - one where the top trapezoidal port meets the vacuum vessel, and second on the inner faces where the stiffeners meet. The maximum deformation is observed
to be around $0.5$ mm at the bottom plate of the vessel where there are DN $40$
ports. At all other surfaces, the stress is less than $40$ MPa. The locations of maximum stress is observed to be mostly at places where the ports (trapezoidal or equatorial) meet the vessel. At these locations, a radius of curvature of $30$ mm is added to relax the local stress concentration. 
 
\begin{figure}[hbt!]
    \centering
        \includegraphics[width=0.8 \linewidth]{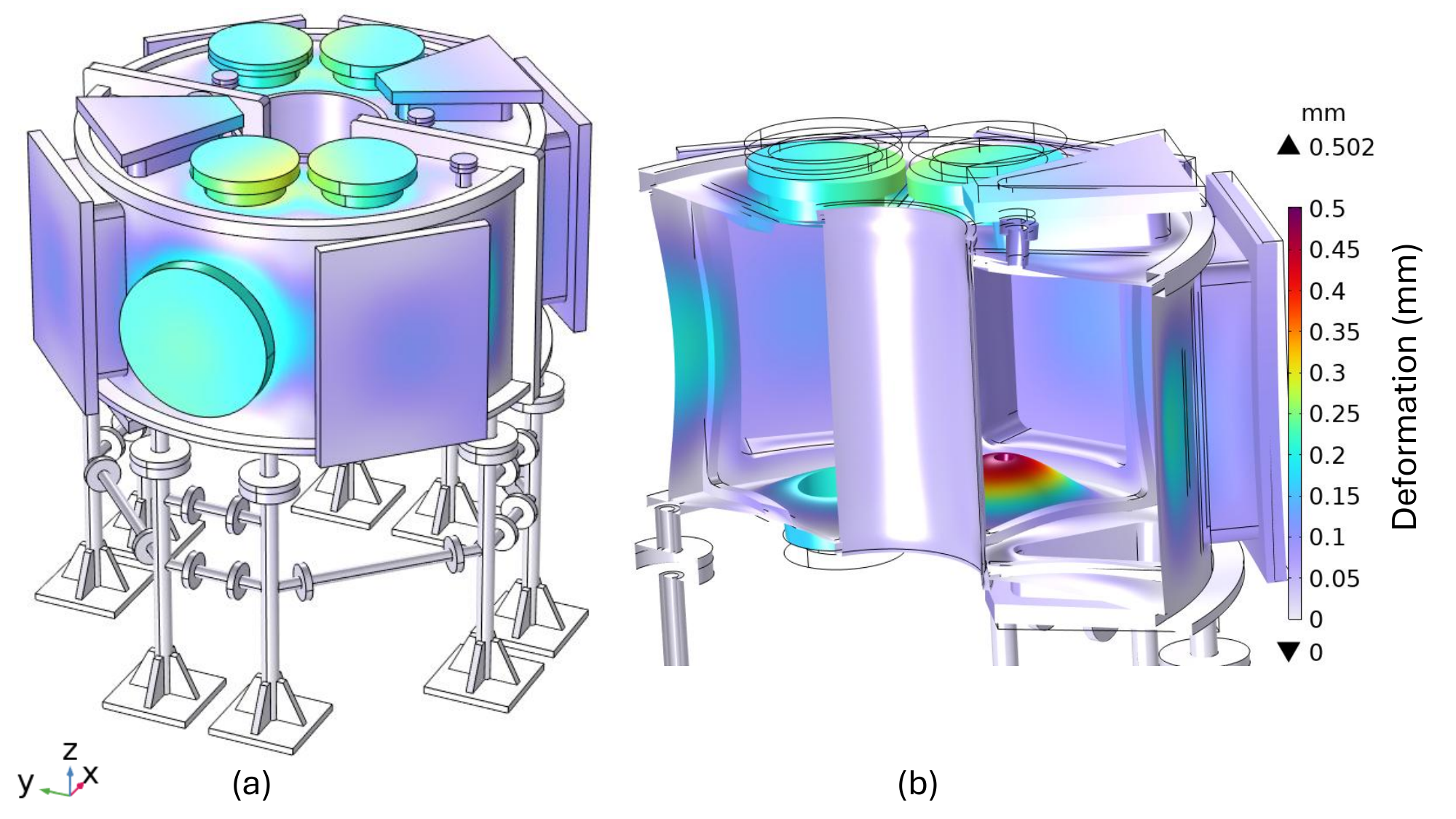}
    \caption{Spatial distribution of deformation (total) on the vacuum vessel due to vacuum and self-weight. In (b), the deformation is scaled $200$ times for the ease of visualization.}
    \label{fig:Pressure_SelfWt_deformation}
\end{figure}

\subsection{Stress due to baking at $150\ ^\circ \mathrm{C}$}
Upon exposure to the ambient environment, $\mathrm{H_2O}$ and
$\mathrm{H_2}$ molecules from the atmosphere get adsorbed in to the steel.
During operations these gases release and contaminate the plasma generated from
the fuel. To minimise the contamination, the vessel is baked at $150$ $^\circ
\mathrm{C}$ for a duration of 48 hours before operations. This removes most of
the absorbed $\mathrm{H_2O}$ gas from the steel. However, heating the vessel to $150$
$^\circ \mathrm{C}$ induces thermal stress which leads to the deformation of the vessel.

\begin{figure}[hbt!]
    \centering
        \includegraphics[width=0.65 \linewidth]{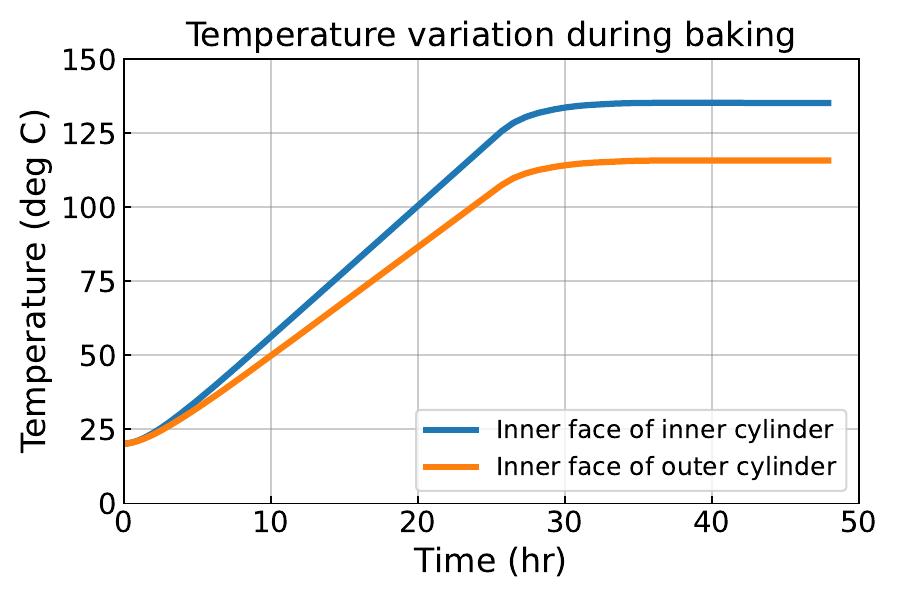}
    \caption{Temporal variation of surface-averaged temperature on the inner face of outer cylinder vacuum vessel due to vacuum, self-weight and baking at $150$ $^\circ \mathrm{C}$.}
    \label{fig:Pressure_thermal_caseQ}
\end{figure}

To assess the effect of baking on the structural integrity of the vacuum vessel,
steady-state and transient thermoelastic (coupled heat transfer and structural
deformation) simulations are performed to identify the regions of high stress and
deformation. Upon fabrication, the \textsc{Pragya} vacuum vessel will be heated
using infrared lamps (IR) on the vacuum facing surfaces. Therefore, in the current
simulations, the vacuum facing surfaces are set to a heat flux boundary
condition such that these surfaces reach a temperature of $150$ $^\circ
\mathrm{C}$. The heat flux is linearly varied with time such that the rate of
temperature increase is limited to $5$ $^\circ \mathrm{C}$ per hour. After
reaching $150$ $^\circ \mathrm{C}$, the input heat flux is kept constant. The
surfaces which are exposed to the atmosphere are set to a convective heat transfer
boundary condition (with heat transfer coefficient, $h = 5\ \mathrm{W/(m^2 \cdot
K)}$) to mimic natural convection. The base of the support legs are set to a
fixed boundary condition. The temporal evolution of the surface-averaged
temperature on the inner face of inner cylinder and inner face of outer cylinder
is shown in Figure \ref{fig:Pressure_thermal_caseQ}. It is observed that the
temperature of these two surfaces increases steadily for $26$ hours after which
it becomes steady. 

\begin{figure}[hbt!]
    \centering
    \includegraphics[width=0.8 \linewidth]{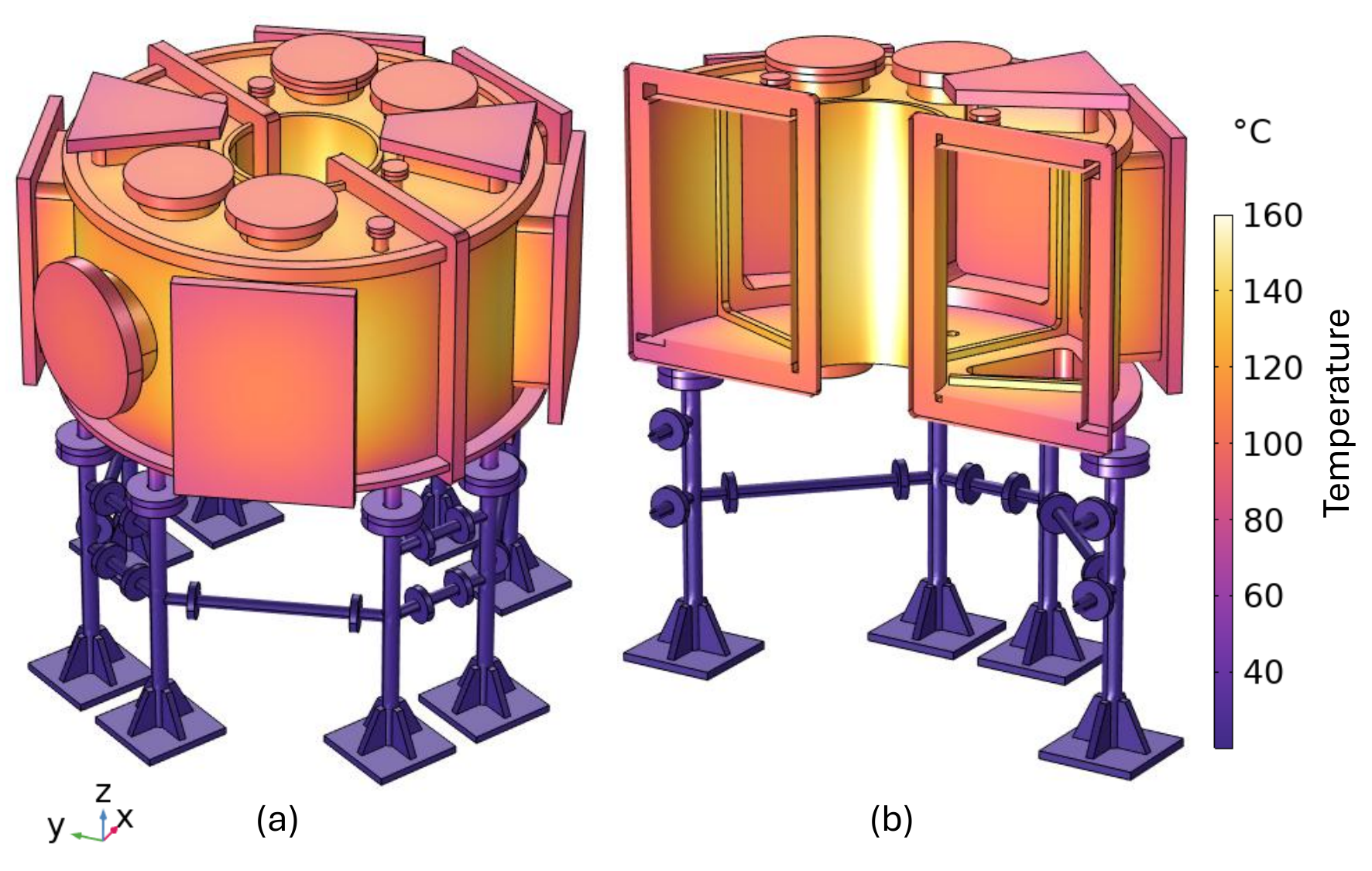}
    \caption{Spatial distribution of temperature on the vacuum vessel due to vacuum, 
             self-weight and baking at $150$ $^\circ \mathrm{C}$.}
    \label{fig:Pressure_thermal_temperature_selfWt}
\end{figure}

\begin{figure}[hbt!]
    \centering
        \includegraphics[width=0.65 \linewidth]{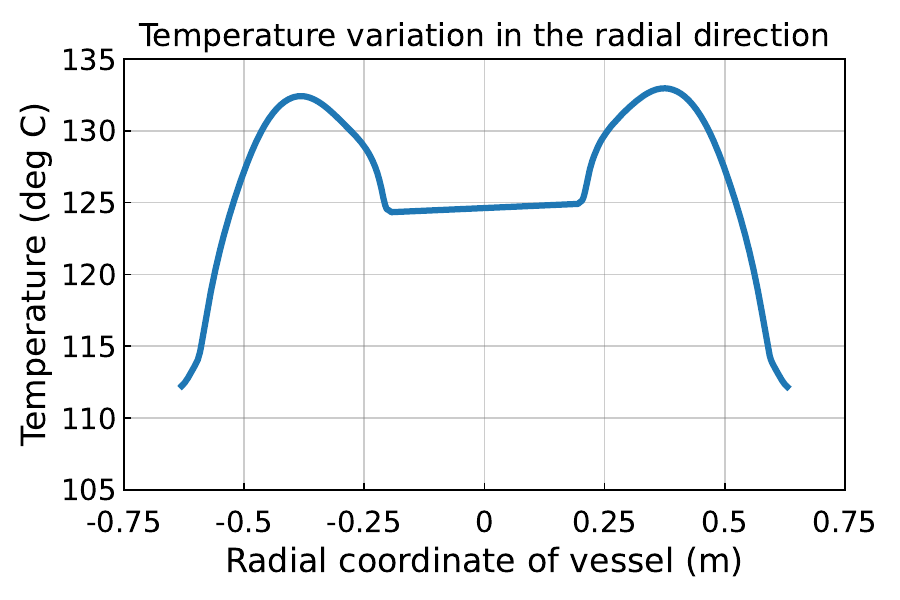}
    \caption{Variation of temperature in the radial direction at the top plate of the vacuum vessel during baking. Coordinate '0' corresponds to the center of the vacuum vessel.}
    \label{fig:Pressure_thermal_tempVariation_radialDirn}
\end{figure}

The spatial distribution of temperature inside and outside the vacuum vessel is shown in Figure \ref{fig:Pressure_thermal_temperature_selfWt}. Although all the vacuum-facing-surfaces are set to an uniform heat flux boundary condition, the temperature distribution is not uniform inside (or outside) the vacuum vessel. While at some spatial location, the temperature does reach (or exceeds) the desired baking temperature i.e. $150$ $^\circ \mathrm{C}$, at other spatial location, the temperature lies between $100$ and $140$ $^\circ \mathrm{C}$. Figure \ref{fig:Pressure_thermal_tempVariation_radialDirn} shows the variation of temperature in the radial direction along the top plate of the vacuum vessel from the inboard side to the outboard side. Since the vacuum vessel is baked at the inner vacuum-facing-surfaces and there is a convective heat transfer boundary condition applied on the outer faces of inboard and outboard sides of the vacuum vessel, a slight drop in temperature is observed as we move away from plasma center to either side (inboard/outboard) of the vacuum vessel. 

\begin{figure}[hbt!]
    \centering
        \includegraphics[width=0.8 \linewidth]{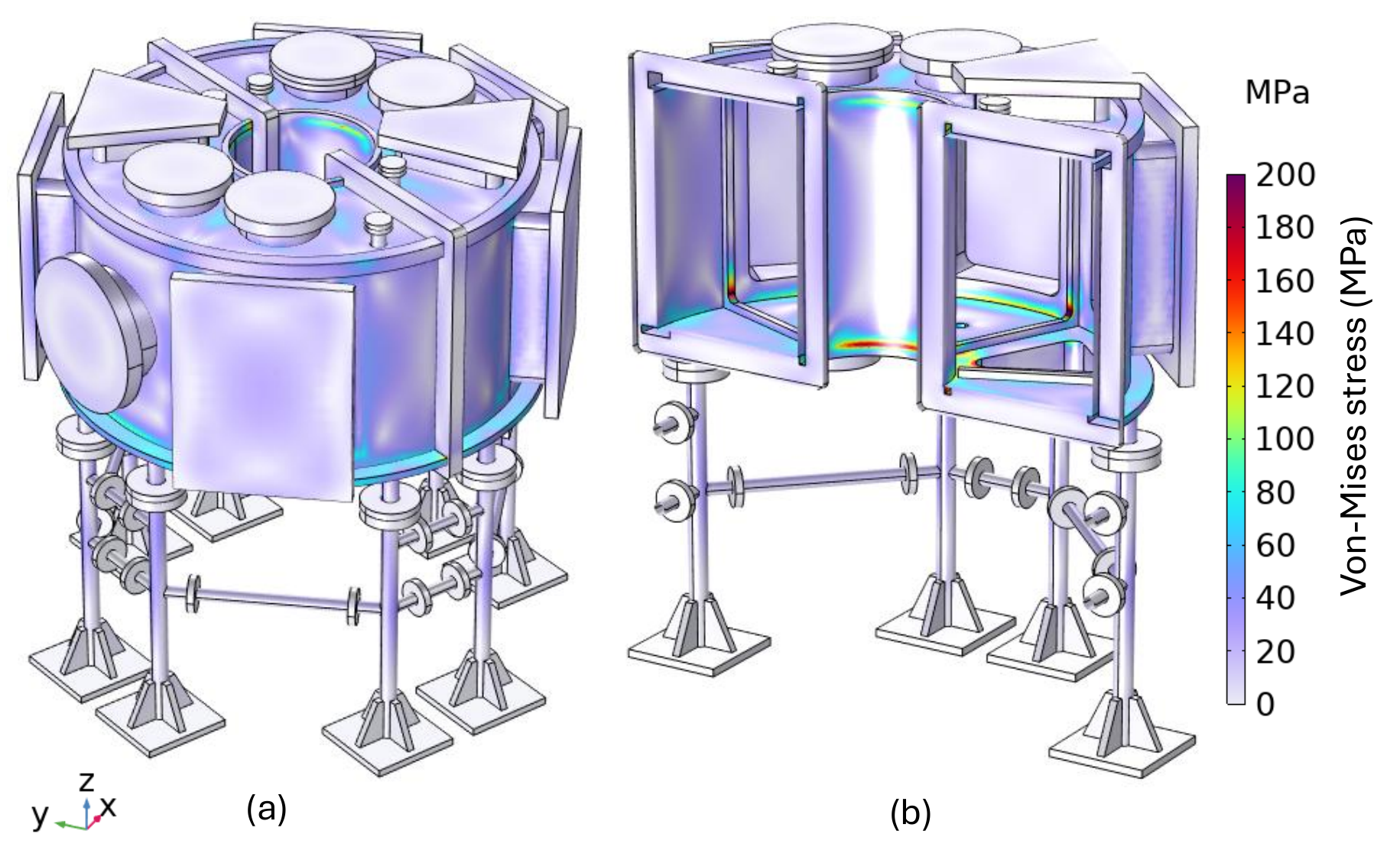}
    \caption{Spatial distribution of stress (Von Mises) on the vacuum vessel due to vacuum, 
             self-weight and baking at $150$ $^\circ \mathrm{C}$.}
    \label{fig:Pressure_thermal_von_mises}
\end{figure}

The stress induced on the vacuum vessel due to simultaneous vacuum and thermal load is shown in Figure \ref{fig:Pressure_thermal_von_mises}. The maximum stress induced is approximately $280$ MPa at two locations - one, on the inboard side of the torus; and second, where the stiffener merges with the inner face of the vacuum vessel as shown in Figure \ref{fig:Pressure_thermal_von_mises}b. However, since the thermal stress is considered as a secondary stress and not a primary stress, the acceptable stress values for the combined primary and secondary stress is twice the yield strength \cite{santra2009thermal, tao2006structural}. Thus, according to ASME criteria, the maximum stress values are within the acceptable limits. On the rest of the vacuum vessel, the thermal stress is less than $80$ MPa.   

\begin{figure}[hbt!]
    \centering
        \includegraphics[width=0.8 \linewidth]{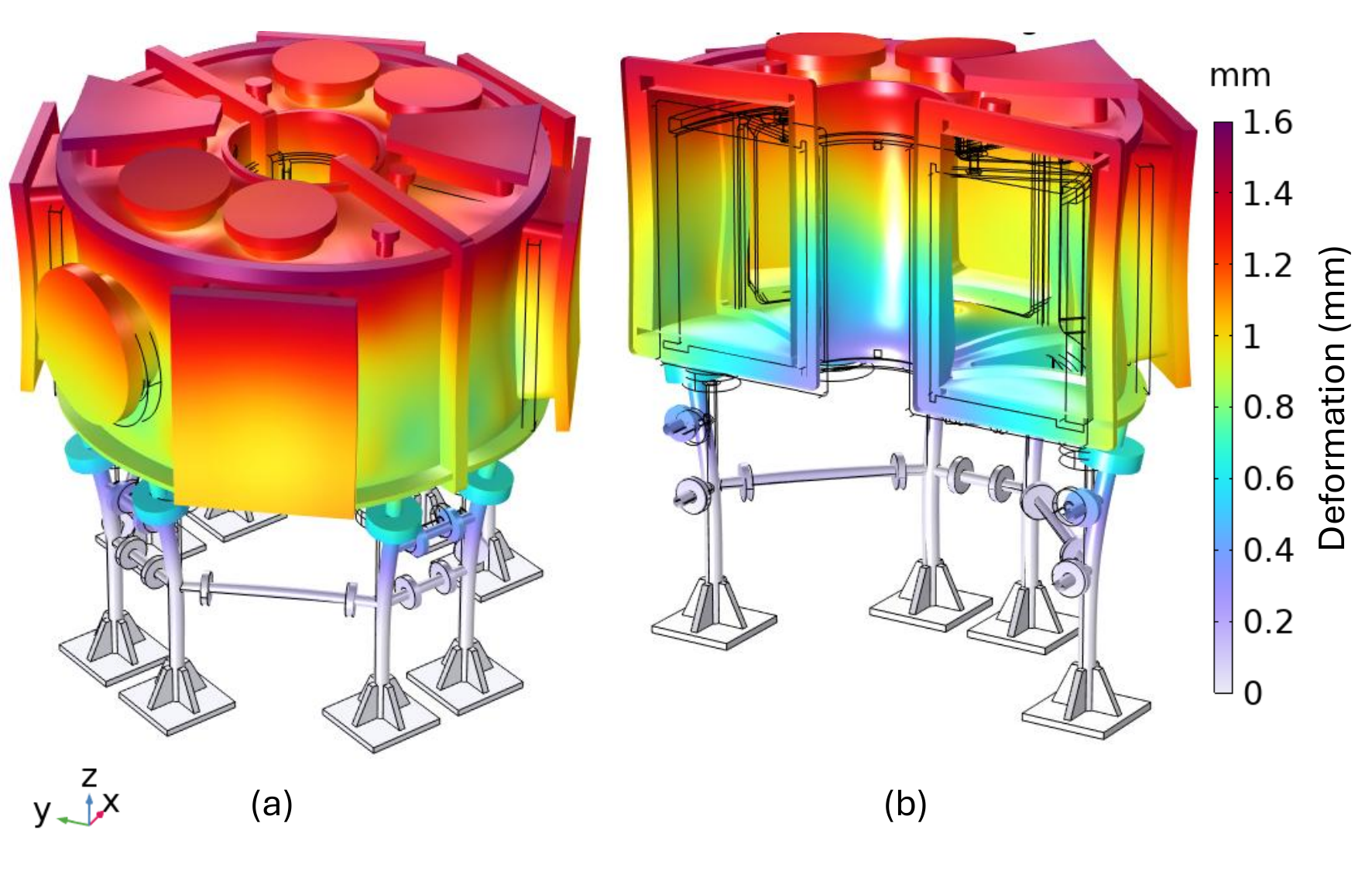}
    \caption{Spatial distribution of deformation (total) on the vacuum vessel due to vacuum, 
             self-weight and baking at $150$ $^\circ \mathrm{C}$. In (b), 
             the deformation is scaled $100$ times for the ease of visualization.}
    \label{fig:Pressure_thermal_deformation}
\end{figure}

Figure \ref{fig:Pressure_thermal_deformation} shows the spatial variation of deformation of the vacuum vessel assembly. Since, the base of the support leg is set to a fixed boundary condition, minimal thermal deformation is observed on the parts of the vacuum vessel close to the support leg, while maximum thermal deformation is observed at locations farthest from the support leg. The maximum deformation is observed to be around $2.1$ mm at the top of the vacuum vessel.

Although, in the simulations the base of support leg is set to a fixed boundary condition, practically the vessel is not fixed to the base of the floor and has provision for slight movement and expansion. The support legs of the vacuum vessel will rest on larger flat base plates (to distribute the load on to larger surface area) which will be grouted to the floor of the laboratory. The support leg will be held onto the flat base plates using bolt arrangement using expandable washers, and larger bolt hole size ($1$ mm larger) than the bolt nominal diameter. These characteristics will provide a flexibility to the support legs for slight movement, which will further reduce the induced thermal stress. The central flange that holds the two sub-tori will also have expandable washers and a larger bolt hole size ($1$ mm larger) to allow for thermal expansion.

\subsection{Linear buckling analysis of the vessel and support legs}
The vessel and the support legs are analysed for buckling due to different loads (weight and pressure difference). This is done to ensure that the structure does not undergo sudden collapse under varying loads acting on it. A linear buckling analysis is performed in two stages: first, one support leg is checked for buckling due to weight of the vessel acting on it; second, the entire vessel is checked for buckling both due to weight and pressure difference.

\subsubsection{Linear buckling analysis of the support legs} \label{ssec:linearBuckling_supportLeg}
A single support leg is checked for buckling due to the weight of the vessel acting on it. The support leg has a annular cross-section with an ID = $28$ mm, OD = $48.30$ mm and length = $700$ mm. Equation \ref{eqn:Euler_critical_load} gives the critical load on a column that can lead to buckling, 

\begin{equation}
    P_{Cr} = \frac{\pi^2 EI}{(KL)^2}
    \label{eqn:Euler_critical_load}
\end{equation}
where $E$, $I$, $K$ and $L$ are Young's modulus ($200$ GPa), minimum second area
moment of inertia of the cross-section of the column, column effective length
factor, and length of the column, respectively. Here, $K = 2$ considering the
support leg is fixed at one end and free at the other end. Following Equation
\ref{eqn:Euler_critical_load}, the critical load on the support leg comes out to
$23,866$ kg. This provides a factor of safety of $120$. Such a high factor of
safety is considered here to allow the vessel withstand other loads acting on it
during operations which are not considered here such as the non-uniformity in
the placement and load distribution on each support leg, moment acting due to
induced Lorentz forces, or a sudden collapse of any one support leg.  

\begin{figure}[hbt!]
    \centering
    \includegraphics[width=0.7 \linewidth]{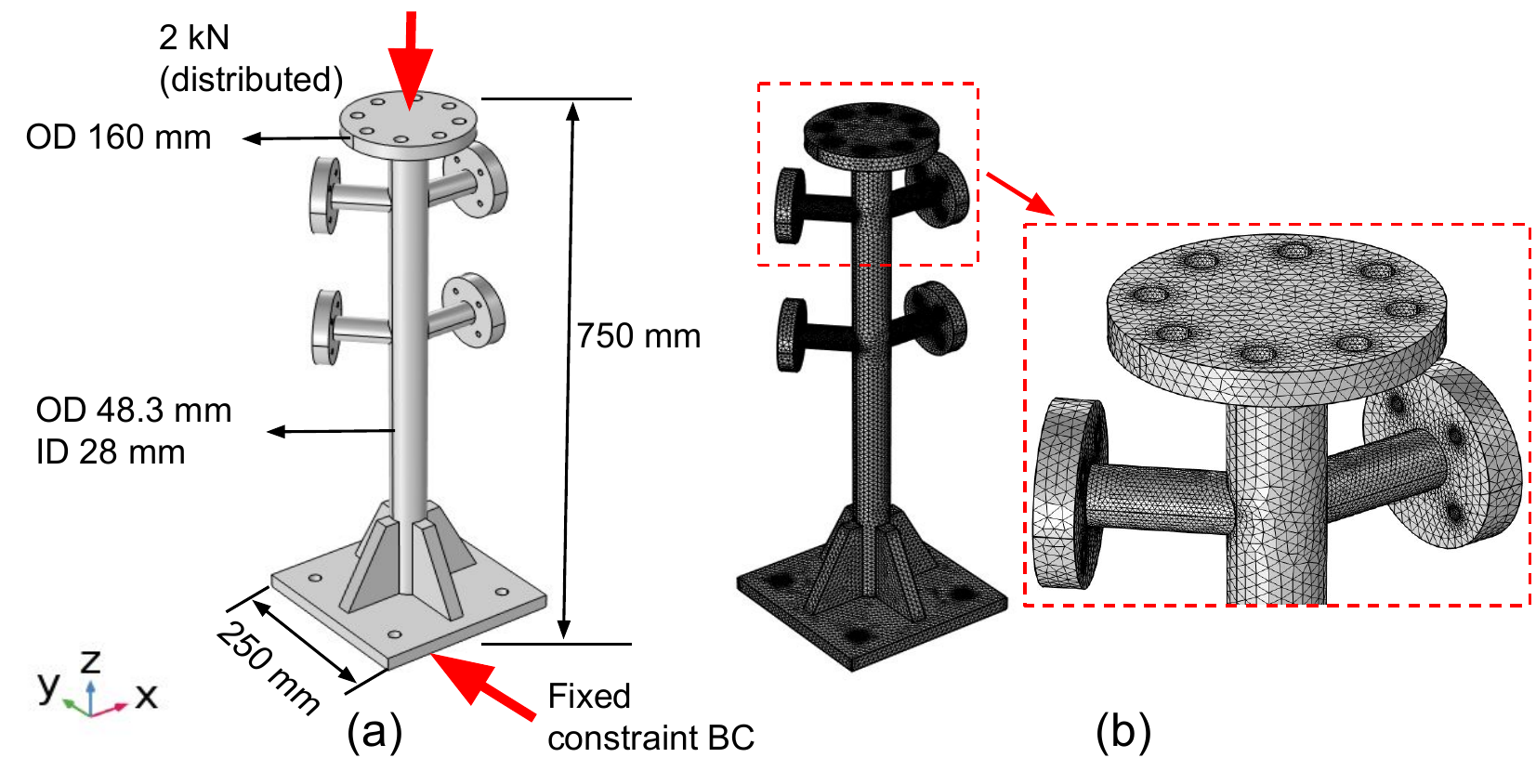}
    \caption{(a) 3D CAD model of the support leg and (b) mesh, during linear buckling analysis.}
    \label{fig:Bucking_supportLeg_CAD_mesh}
\end{figure}

Figure \ref{fig:Bucking_supportLeg_CAD_mesh} shows the 3D CAD, mesh and boundary
conditions acting on the support leg during linear buckling analysis. The base
of the support leg is set to a fixed constraint boundary condition, and a
vertical load of 2 kN is applied on the top face of the support leg.

\begin{figure}[hbt!]
    \centering
    \includegraphics[width=0.9 \linewidth]{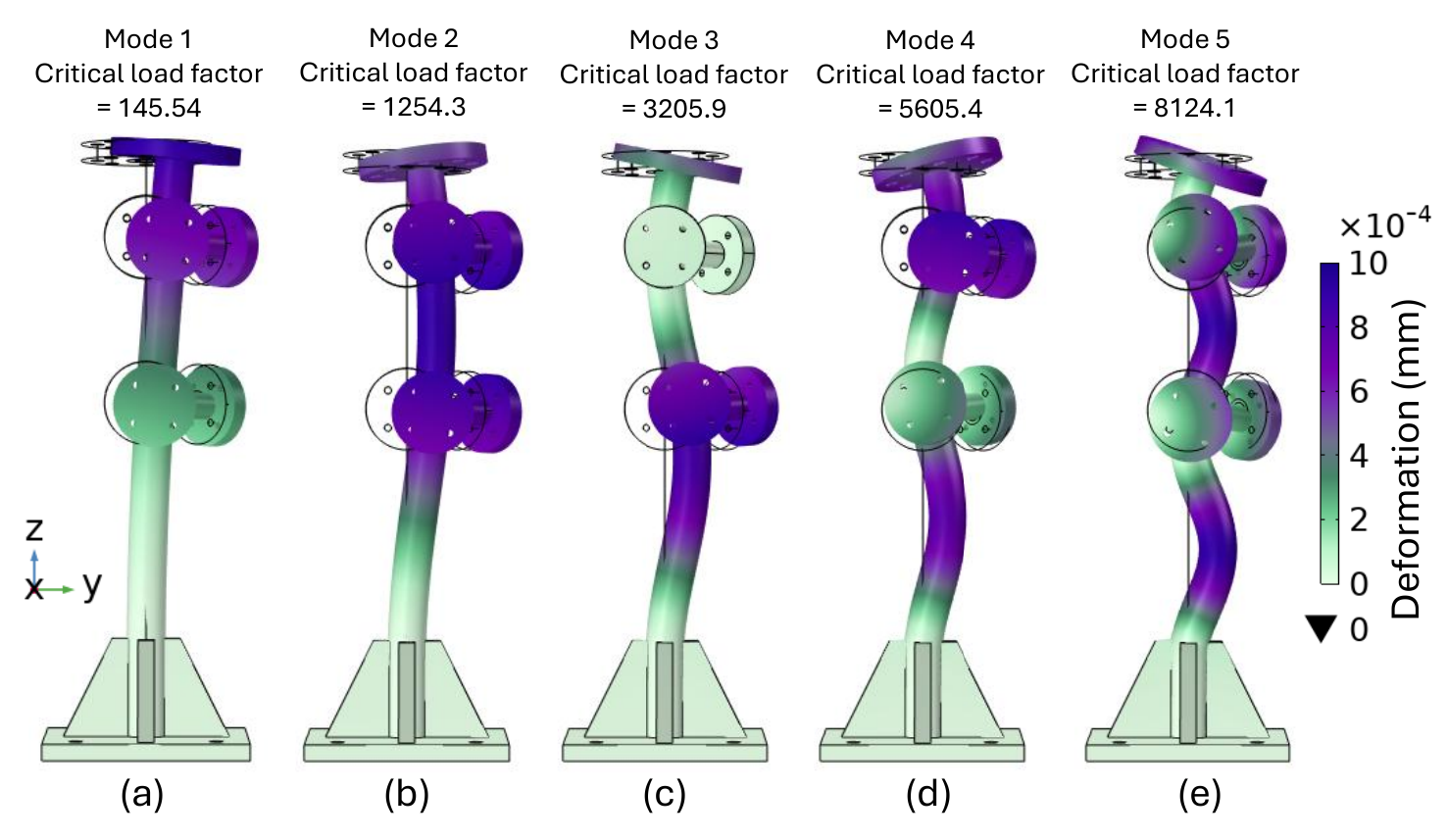}
    \caption{Different modes of deformation and their corresponding critical load factor during vertical load on the support leg (a) Mode 1, (b) Mode 2, (c) Mode 3, (d) Mode 4, and (e) Mode 5.}
    \label{fig:Bucking_supportLeg_modeShape}
\end{figure}

Figure \ref{fig:Bucking_supportLeg_modeShape} shows the different modes of
deformation (or failure) and their corresponding critical load factor (CLF) during vertical load (weight of the vessel) on the support leg.  It is observed that the CLF for all modes is greater than 100. This implies that the required load for the support leg to buckle is at least two orders of magnitude larger than the applied load of $2$ kN, ensuring the support leg is stable during operations. 

\subsubsection{Linear buckling analysis of the vacuum vessel assembly}

\begin{figure}[hbt!]
    \centering
    \includegraphics[width=0.9 \linewidth]{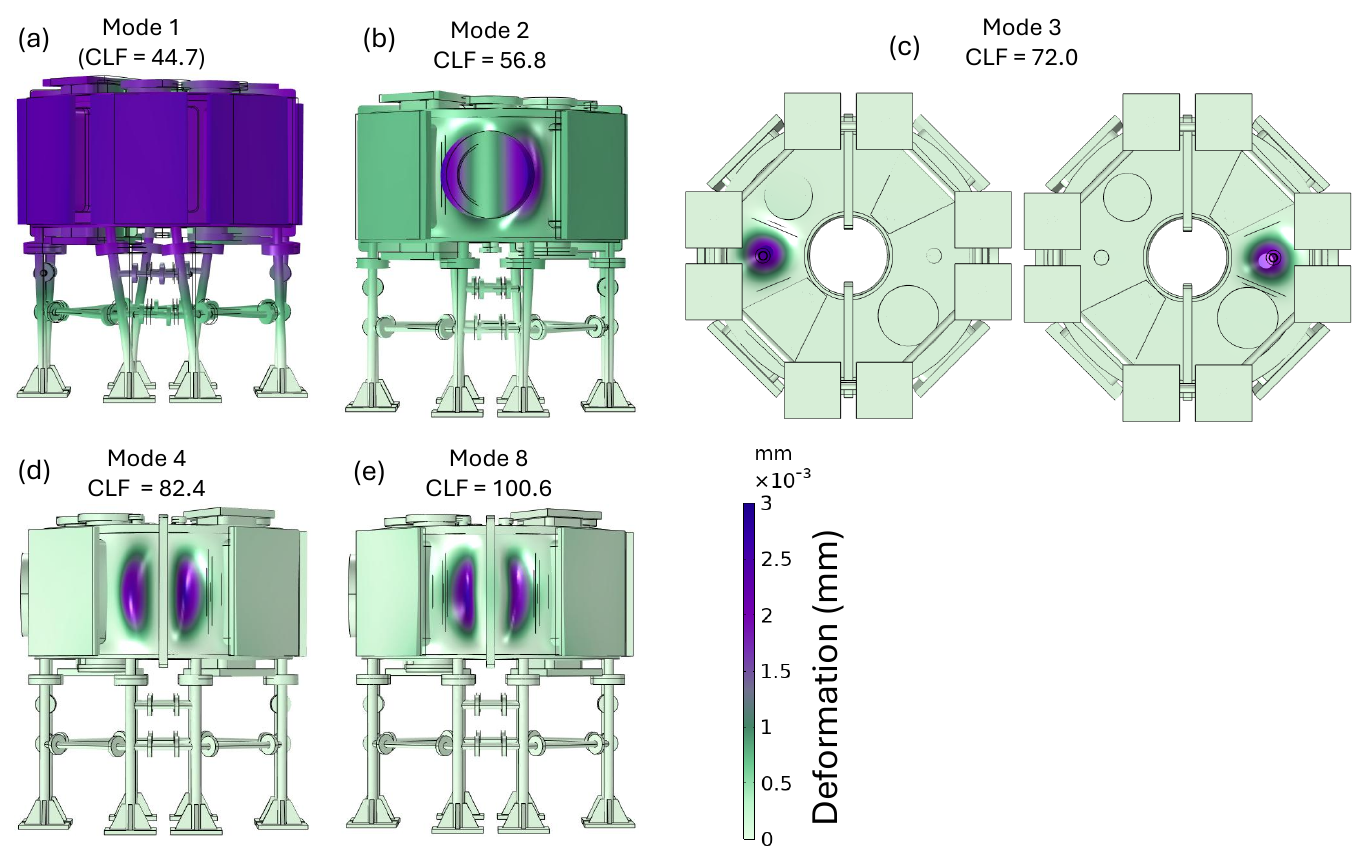}
    \caption{Different modes of deformation and their corresponding critical load factor (CLF) during vertical load on the support leg (a) Mode 1, (b) Mode 2, (c) Mode 3, (d) Mode 4, and (e) Mode 8.}
    \label{fig:Bucking_VV_modeShape}
\end{figure}

In this section, we discuss different modes of failure and their corresponding critical load factors of the entire vacuum vessel assembly due to self-weight and pressure difference. A pressure boundary condition (P = $10^{-5}$ Pa) is applied to all the vacuum-facing-surfaces and atmospheric pressure (P = 101325 Pa) is applied to all the surfaces exposed to atmosphere. The base of the support leg is set to a fixed boundary condition and the gravity acts along the negative $z$ direction. Figure \ref{fig:Bucking_VV_modeShape} shows the different modes of failure and their corresponding CLF. It is observed that there are different modes through which the vacuum vessel can fail such as twisting of the vacuum vessel (mode 1, figure \ref{fig:Bucking_VV_modeShape}a), collapse (or large deformation) around the circular viewing port (mode 2, Figure \ref{fig:Bucking_VV_modeShape}b), large deformation at the bottom face of the vacuum vessel where there are two small DN $40$ ports (mode 3, Figure \ref{fig:Bucking_VV_modeShape}c), or collapse across the central flange connecting the two torus (mode 4 and 8, Figure \ref{fig:Bucking_VV_modeShape}d, \ref{fig:Bucking_VV_modeShape}e). However, the CLF for all these modes is $\gg 1$ which ensures the stability of the vacuum vessel against the applied load.

The linear buckling analysis of the individual support leg and the vacuum vessel assembly shows that they are stable to buckling during self-weight and pressure difference load.


\section{Conclusions} \label{sec:conclusions}
The paper presents the design of the vacuum vessel assembly of the low aspect ratio tokamak- \textsc{Pragya} designed by Pranos Fusion Energy, based in Bangalore, India. The design primarily focuses on the structural and thermal loads acting on the vacuum vessel. \textsc{Pragya} tokamak is designed for a plasma major radius of 0.4 m, a plasma minor radius > 0.18 m and a plasma current of up to 25 kA. The \textsc{Pragya} vacuum vessel has a few distinctive features such as - presence of multiple ports, presence of a toroidal electric break to minimize the induced eddy current on the vacuum vessel, a double O-ring arrangement to minimize the vacuum leak, and stiffeners on the inner face of the vacuum vessel which provide strength to the vessel and have facility to house limiters. Structural, thermal, and linear buckling analysis of the vacuum vessel assembly is performed to identify the load bearing capacity of the vacuum vessel against pressure differential, self-weight and thermal load (due to baking at $150$ $^\circ \mathrm{C}$). It is observed that the stress induced on the vacuum vessel due to pressure and thermal load are within the acceptable limit (less than the yield strength of SS304 L). The maximum stress developed on the vessel is around 110 MPa due to vacuum and self-weight load; with the maximum deformation of $0.5$ mm observed on the bottom side of the vacuum vessel. During baking, a maximum stress of $280$ MPa is observed on the inboard side of the vacuum vessel. Considering thermal stress as secondary stress, the maximum stress is within the acceptable limit. The linear buckling analysis of the individual support leg and the vacuum vessel assembly shows that they are stable to buckling during self-weight and pressure difference load. Future work will consider the effects of electromagnetic load during normal and disruptive events on the vacuum vessel.

\appendix
\renewcommand{\thefigure}{A.\arabic{figure}}
\setcounter{figure}{0} 

\section{Vacuum vessel wall thickness optimization - Geometry and Mesh} \label{Appn:A}

\begin{figure}[hbt!]
    \centering
        \includegraphics[width=0.8 \linewidth]{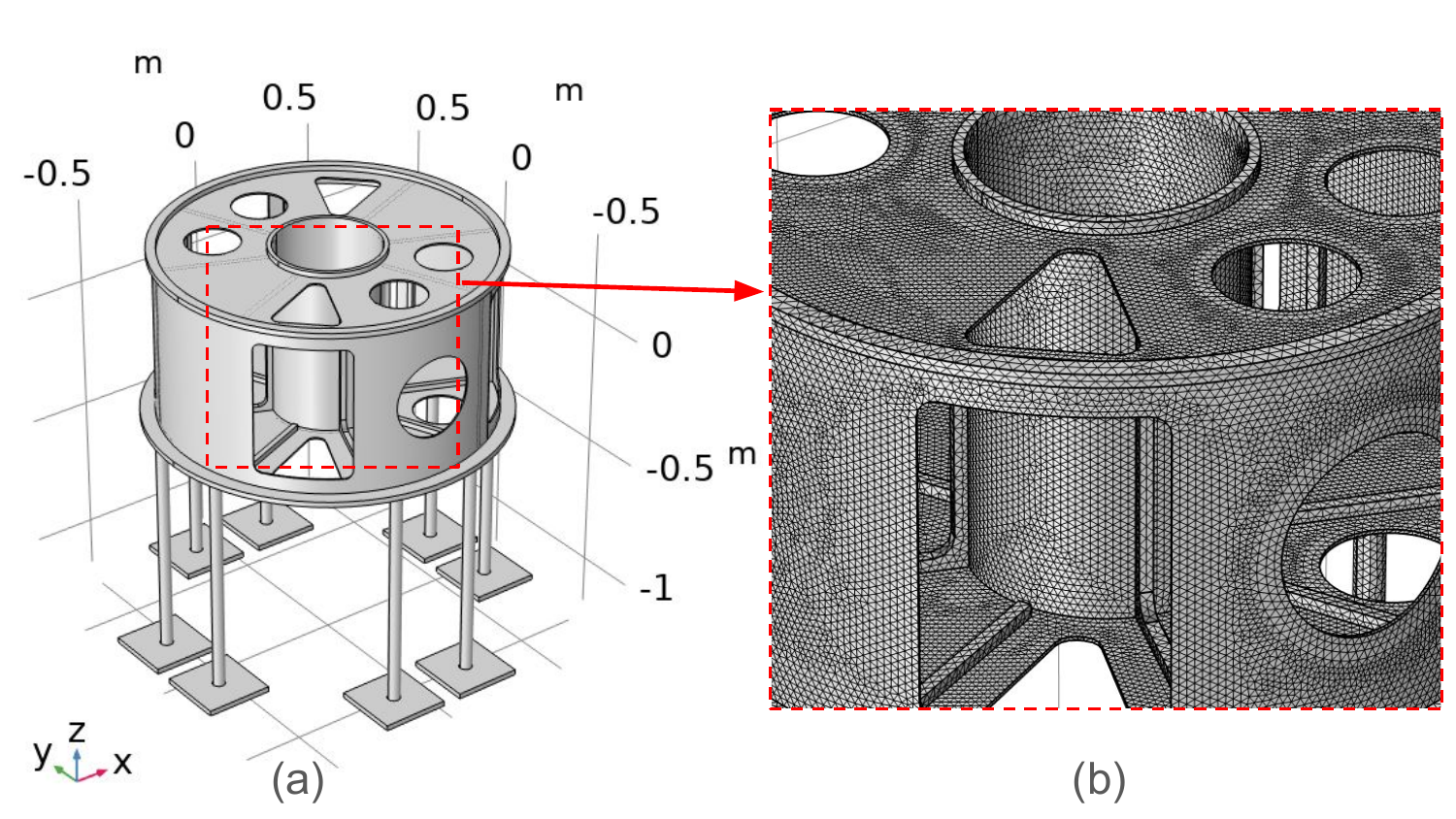}
        \caption{Geometry and mesh used for the wall thickness optimization study.}
        \label{fig:Appn_Geo_Mesh_wallThicknessOptimize}
\end{figure}

This section shows the geometry and the corresponding mesh used during the wall thickness optimization study discussed in Section \ref{sec:stress_analysis}. In the geometry used for optimization study, in addition to the toroidal vacuum vessel, the slots for the ports are provided on different faces of the vacuum vessel. In addition, stiffeners were added on the inside of the vessel. Eight stiffeners are added on each of the following surfaces - top plate, bottom plate and outer cylinder. These stiffeners are present on the inner faces. Four stiffeners are present on the inner face of the inner cylinder. More details on the stiffener can be found in the Section \ref{Sec:MechDesign_Stiffener}. During the wall thickness optimization, the geometry is kept same, only the wall thickness is varied. Figure \ref{fig:Appn_Geo_Mesh_wallThicknessOptimize} shows the geometry and corresponding mesh for case: $6$ mm wall thickness.

\renewcommand{\thefigure}{B.\arabic{figure}}
\setcounter{figure}{0} 

\section{Mechanical analysis with and without stiffeners} \label{Appn:B}

\begin{figure}[hbt!]
    \centering
        \includegraphics[width=0.8 \linewidth]{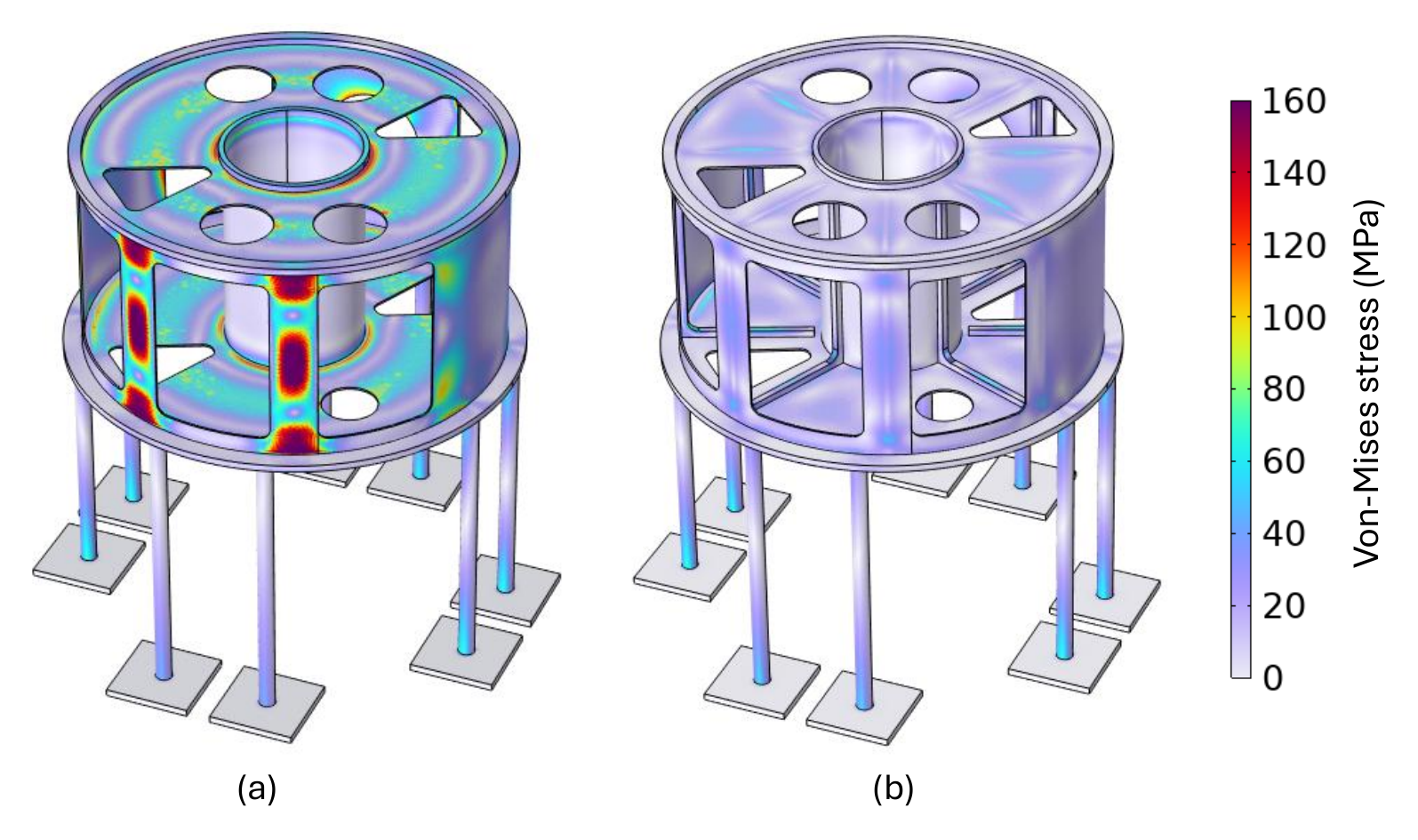}
    \caption{Comparison of von-Mises stress induced on the vacuum vessel (a) without stiffeners, and (b) with stiffeners; Case = $6$ mm wall thickness.}
    \label{fig:Appn:WithAndWithoutStiffener_von_mises_stress}
\end{figure}

This section shows a comparison between the mechanical response of the vacuum vessel with and without the stiffeners. The comparison is made for the vessel wall thickness of $6$ mm. It is observed that adding stiffeners to different faces of vessel results in significant reduction in the observed values of maximum stress. The maximum stress on the cylindrical face of the vacuum vessel reduces from approximately $400$ MPa to $60$ MPa. This is shown in Figure \ref{fig:Appn:WithAndWithoutStiffener_von_mises_stress}. Thus, it is observed that stiffeners are useful in reducing stress levels in the vacuum vessel. Without stiffeners, a thicker vessel will be required to handle such large stress values on the vacuum vessel

\renewcommand{\thefigure}{C.\arabic{figure}}
\setcounter{figure}{0} 

\section{Governing Equations} \label{Appn:C}
The governing equations corresponding to the structural\footnote{\href{https://doc.comsol.com/6.3/doc/com.comsol.help.sme/StructuralMechanicsModuleUsersGuide.pdf}{COMSOL 6.3 Structural Mechanics Module User's Guide}} and themomechanical\footnote{\href{https://doc.comsol.com/6.3/doc/com.comsol.help.heat/HeatTransferModuleUsersGuide.pdf}{COMSOL 6.3 Heat Transfer Module User's Guide}} analysis of the vacuum vessel in COMSOL Multiphysics 6.3 and the fixed-boundary equilibrium in TokaMaker \citep{hansen2024109111} are described in this appendix.

\paragraph{\textbf{Solid deformation}}
The equation of motion of the solid body displacement $\mathbf{u}$ in terms of the Cauchy stress tensor $\boldsymbol{\sigma}$ is given by
\begin{equation}
    \rho \frac{\partial^2 \mathbf{u}}{\partial t^2} = \nabla \cdot \boldsymbol{\sigma} + \mathbf{f}_V
\end{equation}
where $\rho$ is the density and $\mathbf{f_V}$ is the external volume force vector of the deformed solid body. For a linear elastic material, the stress tensor is given by
\begin{equation}
    \boldsymbol{\sigma} = \mathbf{C} : \boldsymbol{\epsilon}
\end{equation}
where $\boldsymbol{\epsilon}$ is the strain tensor, $\mathbf{C}$ is the fourth-order elasticity tensor and ":" indicates a contraction operation over two indices. For an isotropic material, $\mathbf{C}$ reduces to a $6 \times 6$ matrix known as an elasticity matrix $\mathbf{D}$ which is given by
\begin{equation}
    \mathbf{D}  = \frac{E}{(1+\nu)(1-2\nu)}
\begin{bmatrix}
    1-\nu & \nu   & \nu   & 0                & 0                & 0                & \\
      \nu & 1-\nu & \nu   & 0                & 0                & 0                & \\
      \nu & \nu   & 1-\nu & 0                & 0                & 0                & \\
      0   & 0     & 0     & \frac{1-2\nu}{2} & 0                & 0                & \\
      0   & 0     & 0     & 0                & \frac{1-2\nu}{2} & 0                & \\
      0   & 0     & 0     & 0                & 0                & \frac{1-2\nu}{2} & 
\end{bmatrix}
\end{equation}
where $E$ is the Young's modulus and $\nu$ is the Poisson's ratio. 

The Cauchy stress can be decomposed in to isotropic and deviatoric stresses where pressure forms the volumetric isotropic part such that 
\begin{equation}
    p = - \frac{1}{2} \text{trace}(\boldsymbol{\sigma})
\end{equation}
and the deviatoric part is defined as
\begin{equation}
    \sigma_d = \sigma + p \mathbf{I}.
\end{equation}
The von Mises stress is computed from the second invariant of $\sigma_d$ such that
\begin{equation}
    \sigma_{\text{mises}} = \sqrt{\frac{3}{2} \sigma_d : \sigma_d}
\end{equation}

\paragraph{\textbf{Heat transfer in solids}}
The heat transfer in solids is given by
\begin{equation}
    \rho C_p \left( \frac{\partial T}{\partial t} + \mathbf{u} \cdot \nabla T\right) + \nabla \cdot \mathbf{q} = - \alpha T : \frac{dS}{dt} + Q
\end{equation}
where $\rho$ is the solid density, $C_p$ is the specific heat capacity at constant pressure, $T$ is the temperature, $\mathbf{u}$ is the translational velocity, $\mathbf{q}$ is the conduction heat flux, $\alpha$ is the coefficient of thermal expansion, $S$ is the second Piola-Kirchoff stress tensor and $Q$ represents additional volumetric heat sources. Following the Fourier's law of heat conduction, $\mathbf{q}=-k \nabla T$ where $k$ is the thermal conductivity of the solid. The first term on the right hand side of the equation is the thermoelastic damping. The Cauchy stress tensor $\sigma$ and $S$ are related by
\begin{equation}
    \boldsymbol{\sigma} = J^{-1} F S F^T
\end{equation}
where the deformation gradient tensor $F=\nabla \mathbf{u} + \mathbf{I}$ and the ratio between the current and initial mass density is given by $J=\det (F)$. In the coupled system, the solid body experiences additional thermal strain $\epsilon_{\mathrm{Th}}$ which is given by
\begin{equation}
    \epsilon_{\mathrm{Th}} = \alpha \Delta T
\end{equation}
where $\alpha$ is the coefficient of thermal expansion and $\Delta T$ is change in temperature of the solid body.
The resulting constitutive relation is now given by
\begin{equation}
    \boldsymbol{\sigma} = \mathbf{C} : (\epsilon - \epsilon_{\mathrm{Th}}).
\end{equation}

\paragraph{\textbf{Grad-Shafranov equation}}
The Grad-Shafranov equation describes a plasma under static hydrodynamic equilibrium wherein the plasma pressure is balanced by external magnetic field. This is a non-linear elliptic partial differential equation derived in axisymmetric cylindrical coordinates and is given by
 \begin{equation}
       R \frac{\partial}{\partial R} \left(\frac{1}{R}
       \frac{\partial \psi}{\partial R} \right) 
       + \frac{\partial^2 \psi}{\partial Z^2} 
       = -\mu_0 R J_{\phi}(R,Z) 
   \end{equation} \label{eq:gs-full}
where $ \psi $ is the poloidal flux function (V.s/rad), $J_{\phi}$ is the toroidal current density ($\mathrm{A/m^2}$), $\mu_0$ is the permeability of free space. The toroidal current density $J_{\phi}$ is given by
\begin{equation}
    J_{\phi} = R \frac{dp}{d \psi}  + \frac{1}{\mu_0 R}F(\psi)\frac{dF}{d \psi}
\end{equation}
where $p$ is the isotropic plasma pressure (Pa) and $F$ is the toroidal field function defined as $F=R B_{\phi}$ (T.m) where $B_{\phi}$ is the toroidal magnetic field (T).

  \bibliographystyle{elsarticle-num-names} 
  \bibliography{cas-refs}

@article{lawson1957some,
  title={Some criteria for a power producing thermonuclear reactor},
  author={Lawson, John D},
  journal={Proceedings of the physical society. Section B},
  volume={70},
  number={1},
  pages={6},
  year={1957},
  publisher={IOP Publishing}
}

@article{lyons2023flexible,
  title={Flexible, integrated modeling of tokamak stability, transport, equilibrium, and pedestal physics},
  author={Lyons, BC and McClenaghan, J and Slendebroek, T and Meneghini, O and Neiser, TF and Smith, SP and Weisberg, DB and Belli, EA and Candy, J and Hanson, JM and others},
  journal={Physics of Plasmas},
  volume={30},
  number={9},
  year={2023},
  publisher={AIP Publishing}
}

@article{staebler2022advances,
  title={Advances in prediction of tokamak experiments with theory-based models},
  author={Staebler, Gary M and Knolker, M and Snyder, P and Angioni, Clemente and Fable, Emiliano and Luda, T and Bourdelle, Clarisse and Garcia, Jeronimo and Citrin, Jonathan and Marin, Michele and others},
  journal={Nuclear Fusion},
  volume={62},
  number={4},
  pages={042005},
  year={2022},
  publisher={IOP Publishing}
}

@article{whyte2023experimental,
  title={Experimental assessment and model validation of the {SPARC} toroidal field model coil},
  author={Whyte, DG and LaBombard, B and Doody, J and Golfinopolous, T and Granetz, R and Lammi, C and Lane-Walsh, S and Michael, P and Mouratidis, T and Mumgaard, R and others},
  journal={IEEE Transactions on Applied Superconductivity},
  volume={34},
  number={2},
  pages={1--18},
  year={2023},
  publisher={IEEE}
}

@article{hartwig2023sparc,
  title={The {SPARC} toroidal field model coil program},
  author={Hartwig, Zachary S and Vieira, Rui F and Dunn, Darby and Golfinopoulos, Theodore and LaBombard, Brian and Lammi, Christopher J and Michael, Philip C and Agabian, Susan and Arsenault, David and Barnett, Raheem and others},
  journal={IEEE Transactions on Applied Superconductivity},
  volume={34},
  number={2},
  pages={1--16},
  year={2023},
  publisher={IEEE}
}

@article{herrmann2002overview,
doi = {10.1088/0741-3335/44/6/318},
url = {https://doi.org/10.1088/0741-3335/44/6/318},
year = {2002},
month = {may},
publisher = {},
volume = {44},
number = {6},
pages = {883},
author = {A Herrmann},
title = {Overview on stationary and transient divertor heat loads},
journal = {Plasma Physics and Controlled Fusion}
}

@article{federici2001plasma,
  title={Plasma-material interactions in current tokamaks and their implications for next step fusion reactors},
  author={Federici, Gianfranco and Skinner, Charles H and Brooks, Jeffrey N and Coad, Joseph Paul and Grisolia, Christian and Haasz, Anthony A and Hassanein, Ahmed and Philipps, Volker and Pitcher, C Spencer and Roth, Joachim and others},
  journal={Nuclear Fusion},
  volume={41},
  number={12},
  pages={1967},
  year={2001},
  publisher={IOP Publishing}
}

@article{horton1999drift,
  title = {Drift waves and transport},
  author = {Horton, W.},
  journal = {Rev. Mod. Phys.},
  volume = {71},
  issue = {3},
  pages = {735--778},
  numpages = {0},
  year = {1999},
  month = {Apr},
  publisher = {American Physical Society},
  doi = {10.1103/RevModPhys.71.735},
  url = {https://link.aps.org/doi/10.1103/RevModPhys.71.735}
}

@article{hender2007mhd,
  title={{MHD} stability, operational limits and disruptions},
  author={Hender, TC and Wesley, JC and Bialek, J and Bondeson, A and Boozer, AH and Buttery, RJ and Garofalo, A and Goodman, TP and Granetz, RS and Gribov, Y and others},
  journal={Nuclear fusion},
  volume={47},
  number={6},
  pages={S128},
  year={2007},
  publisher={IOP Publishing}
}

@article{ITER_Physics_Expert_Group_on_Confinement_and_Transport_1999,
doi = {10.1088/0029-5515/39/12/302},
url = {https://doi.org/10.1088/0029-5515/39/12/302},
year = {1999},
month = {dec},
publisher = {},
volume = {39},
number = {12},
pages = {2175},
author = {{ITER Physics Expert Group on Confinement and Transport and ITER Physics Expert Group on Confinement Modelling and Database and ITER Physics Basis Editors}},
title = {Chapter 2: Plasma confinement and transport},
journal = {Nuclear Fusion}
}

@article{peng2000physics,
  title={The physics of spherical torus plasmas},
  author={Peng, Y-KM},
  journal={Physics of Plasmas},
  volume={7},
  number={5},
  pages={1681--1692},
  year={2000},
  publisher={American Institute of Physics}
}

@article{ono2015recent,
  title={Recent progress on spherical torus research},
  author={Ono, Masayuki and Kaita, Robert},
  journal={Physics of Plasmas},
  volume={22},
  number={4},
  year={2015},
  publisher={AIP Publishing}
}

@article{mardani2012design,
  title={Design and construction of {A}lborz tokamak vacuum vessel system},
  author={Mardani, M and Amrollahi, R and Koohestani, S},
  journal={Fusion Engineering and Design},
  volume={87},
  number={9},
  pages={1616--1620},
  year={2012},
  publisher={Elsevier}
}

@article{huang2024electromagnetic,
  title={Electromagnetic and mechanical analyses for the vacuum vessel of {NCST}},
  author={Huang, FH and Chen, XC and Chen, H and Liu, SQ},
  journal={Fusion Engineering and Design},
  volume={200},
  pages={114185},
  year={2024},
  publisher={Elsevier}
}

@article{chung2013design,
  title={Design features and commissioning of the {V}ersatile {E}xperiment {S}pherical {T}orus ({VEST}) at {S}eoul {N}ational {U}niversity},
  author={Chung, KJ and An, YH and Jung, BK and Lee, HY and Sung, Choongki and Na, YS and Hahm, TS and Hwang, YS},
  journal={Plasma Science and Technology},
  volume={15},
  number={3},
  pages={244},
  year={2013},
  publisher={IOP Publishing}
}

@article{mancini2021mechanical,
  title={Mechanical and electromagnetic design of the vacuum vessel of the {SMART} tokamak},
  author={Mancini, Alessio and Ayllon-Guerola, J and Doyle, Scott J and Agredano-Torres, Manuel and Lopez-Aires, D and Toledo-Garrido, J and Viezzer, Eleonora and Garcia-Mu{\~n}oz, M and Buxton, Peter F and Chung, Kyoung-Jae and others},
  journal={Fusion Engineering and Design},
  volume={171},
  pages={112542},
  year={2021},
  publisher={Elsevier}
}

@article{jadeja2017aditya,
  title={{ADITYA} upgrade vacuum vessel: {D}esign, construction, testing, installation and operation},
  author={Jadeja, KA and Bhatt, SB and Rathod, Kulav and Patel, KM and Prajapati, VR and Acharya, KS and Patel, ND and Tanna, RL and Kalal, MB and Ghosh, J and others},
  journal={Fusion Engineering and Design},
  volume={124},
  pages={558--561},
  year={2017},
  publisher={Elsevier}
}

@article{li2015measurement,
  title={Measurement of eddy-current distribution in the vacuum vessel of the {S}ino-{UNI}ted {S}pherical {T}okamak},
  author={Li, G and Tan, Y and Liu, YQ},
  journal={Review of scientific instruments},
  volume={86},
  number={8},
  year={2015},
  publisher={AIP Publishing}
}

@inproceedings{finkelmeyer1979assembly,
  title={Assembly and commissioning of the {ASDEX} tokamak},
  author={Finkelmeyer, H and Franzspeck, J and Gernhardt, J and Gresser, F and Haas, G and Hartz, F and Herppich, G and Keilhacker, M and Klement, G and Kornherr, M},
  booktitle={8th Symposium on Engineering Problems of Fusion Research},
  volume={3},
  pages={1299--1302},
  year={1979}
}

@article{sontag2022new,
  title={The new {PEGASUS-III} experiment},
  author={Sontag, Aaron C and Bongard, Michael W and Borchardt, Michael T and Diem, Stephanie J and Fonck, Raymond J and Keyhani, Armand K and Kujak-Ford, Benjamin A and Lewicki, Benjamin T and Nornberg, Mark D and Palmer, Alan C and others},
  journal={IEEE Transactions on Plasma Science},
  volume={50},
  number={11},
  pages={4009--4014},
  year={2022},
  publisher={IEEE}
}

@article{ying2003initial,
  title={Initial plasma startup test on {SUNIST} spherical tokamak},
  author={Ying, Wang and Li, Zeng and Yexi, He and SUNIST Team and others},
  journal={Plasma Science and Technology},
  volume={5},
  number={6},
  pages={2017},
  year={2003},
  publisher={IOP Publishing}
}

@article{sakurai2009design,
  title={Design and manufacturing of vacuum vessel for {JT-60SA}},
  author={Sakurai, Shinji and Masaki, Kei and Shibama, Yusuke K and Sakasai, Akira},
  journal={Fusion engineering and design},
  volume={84},
  number={7-11},
  pages={1684--1688},
  year={2009},
  publisher={Elsevier}
}

@inproceedings{song2006design,
  title={Design, fabrication and testing results of vacuum vessel, thermal shield and cryostat of {EAST}},
  author={Song, YT and Yao, D and Liao, Z and Yu, J and Xie, H and Wu, W and Gao, D and Wu, S and Li, J and Weng, P and others},
  booktitle={21th International Atomic Energy Agency Fusion Energy Conference in Cheng Du, China 16-21 Oct},
  year={2006}
}

@article{santra2009thermal,
  title={Thermal structural analysis of {SST-1} vacuum vessel and cryostat assembly using {ANSYS}},
  author={Santra, Prosenjit and Bedakihale, Vijay and Ranganath, Tata},
  journal={Fusion engineering and design},
  volume={84},
  number={7-11},
  pages={1708--1712},
  year={2009},
  publisher={Elsevier}
}

@article{tao2006structural,
  title={Structural analysis and manufacture for the vacuum vessel of experimental advanced superconducting tokamak ({EAST}) device},
  author={tao Song, Yun and Yao, Damao and Wu, Songata and Weng, Peide},
  journal={Fusion engineering and design},
  volume={81},
  number={8-14},
  pages={1117--1122},
  year={2006},
  publisher={Elsevier}
}

@article{darke1995MAST,
author="Darke, AC and Harbar, JR and Hay, JH and Hicks, JB and Hill, JW and McKenzie, JS and Morris, AW and Nightingale, MPS and Todd, TN and Voss, GM and Watkins, JR",
title="{MAST}: A {M}ega {A}mp {S}pherical {T}okamak",
journal="Fusion Technology 1994",
publisher="Elsevier",
year="1995",
pages="799-802",
DOI="10.1016/b978-0-444-82220-8.50167-9"
}

@article{ono2001overview,
  title={Overview of the initial {NSTX} experimental results},
  author={Ono, Masayuki and Bell, MG and Bell, RE and Bigelow, T and Bitter, M and Blanchard, W and Darrow, DS and Fredrickson, ED and Gates, DA and Grisham, LR and others},
  journal={Nuclear fusion},
  volume={41},
  number={10},
  pages={1435},
  year={2001},
  publisher={IOP Publishing}
}

@article{keilhacker1999jet,
  title={{JET} deuterium: tritium results and their implications},
  author={Keilhacker, M},
  journal={Philosophical Transactions of the Royal Society of London. Series A: Mathematical, Physical and Engineering Sciences},
  volume={357},
  number={1752},
  pages={415--442},
  year={1999},
  publisher={The Royal Society}
}

@article{strachan1997tftr,
  title={{TFTR} {DT} experiments},
  author={Strachan, JD and Batha, S and Beer, M and Bell, MG and Bell, RE and Belov, A and Berk, H and Bernabei, S and Bitter, M and Breizman, B and others},
  journal={Plasma Physics and Controlled Fusion},
  volume={39},
  number={12B},
  pages={B103},
  year={1997},
  publisher={IOP Publishing}
}

@article{kishimoto2005advanced,
  title={Advanced tokamak research on {JT}-60},
  author={Kishimoto, H and Ishida, S and Kikuchi, M and Ninomiya, H},
  journal={Nuclear fusion},
  volume={45},
  number={8},
  pages={986},
  year={2005},
  publisher={IOP Publishing}
}

@article{wagner1982regime,
  title={Regime of improved confinement and high beta in neutral-beam-heated divertor discharges of the ASDEX tokamak},
  author={Wagner, Fritz and Becker, G and Behringer, K and Campbell, D and Eberhagen, A and Engelhardt, W and Fussmann, G and Gehre, O and Gernhardt, J and Gierke, G v and others},
  journal={Physical Review Letters},
  volume={49},
  number={19},
  pages={1408},
  year={1982},
  publisher={APS}
}

@article{hansen2024109111,
title = {Toka{M}aker: An open-source time-dependent Grad-Shafranov tool for the design and modeling of axisymmetric fusion devices},
journal = {Computer Physics Communications},
volume = {298},
pages = {109111},
year = {2024},
issn = {0010-4655},
doi = {https://doi.org/10.1016/j.cpc.2024.109111},
url = {https://www.sciencedirect.com/science/article/pii/S0010465524000341},
author = {C. Hansen and I.G. Stewart and D. Burgess and M. Pharr and S. Guizzo and F. Logak and A.O. Nelson and C. Paz-Soldan}
}

@article{tomarchio20173,
title = {Status of the {JT-60SA} project: An overview on fabrication, assembly and future exploitation},
journal = {Fusion Engineering and Design},
volume = {123},
pages = {3-10},
year = {2017},
note = {Proceedings of the 29th Symposium on Fusion Technology (SOFT-29) Prague, Czech Republic, September 5-9, 2016},
issn = {0920-3796},
doi = {https://doi.org/10.1016/j.fusengdes.2017.05.041},
url = {https://www.sciencedirect.com/science/article/pii/S0920379617305859},
author = {V. Tomarchio and P. Barabaschi and E. {Di Pietro} and M. Hanada and Y. Kamada and A. Sakasai and H. Shirai}
}

@article{cpcwong_2002,
doi = {10.1088/0029-5515/42/5/307},
url = {https://doi.org/10.1088/0029-5515/42/5/307},
year = {2002},
month = {may},
publisher = {},
volume = {42},
number = {5},
pages = {547},
author = {C.P.C. Wong and J.C. Wesley and R.D. Stambaugh and E.T. Cheng},
title = {Toroidal reactor designs as a function of  aspect 
ratio and elongation},
journal = {Nuclear Fusion}
}

@book{ASME_BPVC_II_D_2019,
  title     = {Boiler and Pressure Vessel Code, Section II---Materials, Part D: Properties},
  author    = {{American Society of Mechanical Engineers}},
  year      = {2019},
  publisher = {ASME},
  address   = {New York, NY}
}

@conference{millerstable1996,
  author       = {Miller, R L and Lin-Liu, Y R and Turnbull, A D and Chan, V S and Pearlstein, L D and Sauter, O and Villard, L},
  title        = {Stable bootstrap-current driven equilibria for low aspect ratio tokamaks},
  url          = {https://www.osti.gov/biblio/390606},
  place        = {United States},
  organization = {General Atomics, San Diego, CA (United States); Lawrence Berkeley Lab., CA (United States)},
  year         = {1996},
  month        = {08}}

@article{wu2007overview,
  title={An overview of the {EAST} project},
  author={Wu, Songtao and {EAST} team and others},
  journal={Fusion Engineering and Design},
  volume={82},
  number={5-14},
  pages={463--471},
  year={2007},
  publisher={Elsevier}
}

@inproceedings{deshpande1997sst,
  title={{SST}-1: an overview},
  author={Deshpande, SP},
  booktitle={17th IEEE/NPSS Symposium Fusion Engineering (Cat. No. 97CH36131)},
  volume={1},
  pages={227--232},
  year={1997},
  organization={IEEE}
}

@article{bucalossi2022operating,
  title={Operating a full tungsten actively cooled tokamak: overview of WEST first phase of operation},
  author={Bucalossi, J and Achard, J and Agullo, O and Alarcon, T and Allegretti, L and Ancher, H and Antar, G and Antusch, S and Anzallo, V and Arnas, C and others},
  journal={Nuclear Fusion},
  volume={62},
  number={4},
  pages={042007},
  year={2022},
  publisher={IOP Publishing}
}

@article{luxon2002design,
  title={A design retrospective of the {DIII-D} tokamak},
  author={Luxon, James L},
  journal={Nuclear Fusion},
  volume={42},
  number={5},
  pages={614},
  year={2002},
  publisher={IOP Publishing}
}

@article{lee2001design,
  title={Design and construction of the {KSTAR} tokamak},
  author={Lee, GS and Kwon, M and Doh, CJ and Hong, BG and Kim, K and Cho, MH and Namkung, W and Chang, Choong-Seock and Kim, YC and Kim, JY and others},
  journal={Nuclear Fusion},
  volume={41},
  number={10},
  pages={1515},
  year={2001},
  publisher={IOP Publishing}
}

@article{creely2020overview,
  title={Overview of the {SPARC} tokamak},
  author={Creely, AJ and Greenwald, Martin J and Ballinger, Sean B and Brunner, D and Canik, J and Doody, Jeffrey and F{\"u}l{\"o}p, T and Garnier, DT and Granetz, R and Gray, TK and others},
  journal={Journal of Plasma Physics},
  volume={86},
  number={5},
  pages={865860502},
  year={2020},
  publisher={Cambridge University Press}
}

@article{fukuyama1975positional,
  title={Positional instabilities in a tokamak with a resistive shell},
  author={Fukuyama, Atsushi and Seki, Shogo and Momota, Hiromu and Itatani, Ryohei},
  journal={Japanese Journal of Applied Physics},
  volume={14},
  number={6},
  pages={871},
  year={1975},
  publisher={IOP Publishing}
}

@article{maekawa2005formation,
  title={Formation of spherical tokamak equilibria by {ECH} in the {LATE} device},
  author={Maekawa, T and Terumichi, Y and Tanaka, H and Uchida, M and Yoshinaga, T and Yamaguchi, S and Igami, H and Konno, M and Katsuura, K and Hayashi, K and others},
  journal={Nuclear fusion},
  volume={45},
  number={11},
  pages={1439},
  year={2005},
  publisher={IOP Publishing}
}

@article{boyle2023extending,
  title={Extending the low-recycling, flat temperature profile regime in the lithium tokamak experiment-$\beta$ ({LTX}-$\beta$) with ohmic and neutral beam heating},
  author={Boyle, Dennis P and Anderson, J and Banerjee, S and Bell, RE and Capecchi, W and Elliott, DB and Hansen, C and Kubota, S and LeBlanc, BP and Maan, A and others},
  journal={Nuclear Fusion},
  volume={63},
  number={5},
  pages={056020},
  year={2023},
  publisher={IOP Publishing}
}

@article{banerjee2024investigating,
  title={Investigating the role of edge neutrals in exciting tearing mode activity and achieving flat temperature profiles in {LTX}-$\beta$},
  author={Banerjee, Santanu and Boyle, DP and Maan, A and Ferraro, N and Wilkie, G and Majeski, R and Podesta, M and Bell, R and Hansen, C and Capecchi, W and others},
  journal={Nuclear Fusion},
  volume={64},
  number={4},
  pages={046026},
  year={2024},
  publisher={IOP Publishing}
}

@article{marmar2007alcator,
  title={The {A}lcator {C}-{M}od program},
  author={Marmar, Earl S and Group, Alcator C-Mod},
  journal={Fusion science and technology},
  volume={51},
  number={3},
  pages={261--265},
  year={2007},
  publisher={Taylor \& Francis}
}

@article{bhatt_bora_buch_and_others_1989, 
title={Aditya : the first Indian tokamak}, 
volume={27}, 
ISSN={0019-5596}, 
abstractNote={One of the methods of confining hot fusion plasmas is by use of the toroidal magnetic bottle known as tokamak. Aditya is the first tokamak, designed, fabricated, erected and commissioned by India. It is a moderate field (toroidal field 15 kGauss), medium size (minor radius 25 cm, major radius 75 cm) tokamak which should produce a 250 kA discharge with a peak density 3x1013 cm-3 and a peak electron temperature of about 5 million degrees. The plasma is generated in a stainless steel vessel evacuated to a base pressure of 10-9 torr. The vessel is surrounded by twenty large rectangular torroidal magnetic field coils. An ohmic transformer is placed in the central hole of the torus and is used to induce substantial voltage in the hydrogen gas (typically kept at 10-4 torr in the vessel) to cause its breakdown and drive and sustain plasma currents upto 250 kA. Other field coils produce programmed fields necessary for keeping the plasmas in equilibrium. Power is supplied to the various magnetic field coils in the form of dc pulses generated by transformer-converter systems controlled by a PDP 11/23 computer. The main source of the power is a 132 kV line connected to the Gujarat Electricity Board power grid system. Plasma parameters are to be measured by a variety of passive and active diagnostics. The data acquisition and control of the entire experiment is carried out using control system based on CAMAC and VAX 11/730 computer. Experiments dealing with plasma disruptions and their feedback control, plasmas with significant current in energetic carriers, etc. are the major directions of the experimental programme. The paper summarises design and development activity leading to the commissioning of Aditya. (author). 41 refs., 31 figs., 16 tabs}, 
number={9-10}, 
journal={Indian Journal of Pure and Applied Physics}, 
author={Bhatt, S.B. and Bora, D. and Buch, B.N. and and others}, 
year={1989}, 
month={Sep}, 
pages={710–742} 
}

@ARTICLE{kerboua2024,
  author={Kerboua-Benlarbi, S. and Nouailletas, R. and Faugeras, B. and Nardon, E. and Moreau, P.},
  journal={IEEE Transactions on Plasma Science}, 
  title={Magnetic Control of {WEST} Plasmas Through Deep Reinforcement Learning}, 
  year={2024},
  volume={52},
  number={9},
  pages={3698-3703},
  doi={10.1109/TPS.2024.3377811}
}

@Article{Degrave2022,
author={Degrave, Jonas
and Felici, Federico
and Buchli, Jonas
and Neunert, Michael
and Tracey, Brendan
and Carpanese, Francesco
and Ewalds, Timo
and Hafner, Roland
and Abdolmaleki, Abbas
and de las Casas, Diego
and Donner, Craig
and Fritz, Leslie
and Galperti, Cristian
and Huber, Andrea
and Keeling, James
and Tsimpoukelli, Maria
and Kay, Jackie
and Merle, Antoine
and Moret, Jean-Marc
and Noury, Seb
and Pesamosca, Federico
and Pfau, David
and Sauter, Olivier
and Sommariva, Cristian
and Coda, Stefano
and Duval, Basil
and Fasoli, Ambrogio
and Kohli, Pushmeet
and Kavukcuoglu, Koray
and Hassabis, Demis
and Riedmiller, Martin},
title={Magnetic control of tokamak plasmas through deep reinforcement learning},
journal={Nature},
year={2022},
month={Feb},
day={01},
volume={602},
number={7897},
pages={414-419},
issn={1476-4687},
doi={10.1038/s41586-021-04301-9},
url={https://doi.org/10.1038/s41586-021-04301-9}
}

@Article{Jenko2025,
author={Jenko, Frank},
title={Accelerating fusion research via supercomputing},
journal={Nature Reviews Physics},
year={2025},
month={Jul},
day={01},
volume={7},
number={7},
pages={365-377},
issn={2522-5820},
doi={10.1038/s42254-025-00837-1},
url={https://doi.org/10.1038/s42254-025-00837-1}
}






\end{document}